\documentclass[aip, pof, reprint]{revtex4-2}

\usepackage{XiGrpCommon}
\usepackage{subcaption}
\usepackage{amsmath}
\usepackage{graphicx}
\usepackage{comment}

\graphicspath{{figures/}}

\usepackage{CJKutf8}

\renewcommand{\RevisedText}[1]{#1}

\begin{document}
\begin{CJK*}{UTF8}{gbsn}
\CJKtilde 
\CJKindent 

\title{Dynamics and Stress Relaxation of Bidisperse Polymer Melts with Unentangled and Moderately Entangled Chains}
\author{Oluseye Adeyemi}
\author{Shiping Zhu (朱世平)}
\author{Li Xi (奚力)}
\email[coresponding author, E-mail: ]{xili@mcmaster.ca}
\homepage[Web: ]{https://www.xiresearch.org}
\affiliation{Department of Chemical Engineering, McMaster University, Hamilton, Ontario L8S 4L7, Canada}

\newpage
\begin{abstract}
Polydispersity is inevitable in industrially produced polymers. Established theories of polymer dynamics and rheology, however, were mostly built on monodisperse linear polymers. 
Dynamics of polydisperse polymers is yet to be fully explored -- especially how chains of different lengths affect the dynamics of one another in a mixture.
This study explored the dynamics of bidisperse polymer melts using molecular dynamics and a bead-spring chain model. Binary mixtures between a moderately entangled long-chain species and an unentangled or marginally entangled short-chain species were investigated.
We found that adding short chains can significantly accelerate the dynamics of the long chains by substantially lessening their extent of entanglement.
Meanwhile, although introducing long chains also hinders the motion of the short chains, it does not qualitatively alter the nature of their dynamics -- unentangled short chains still follow classical Rouse dynamics even in a matrix containing entangled chains.
Detailed Rouse mode analysis was used to reveal the effects of entanglement at chain segments of different scales.
Stress relaxation following a step shear strain was also studied and semi-empirical mixing rules that predict the linear viscoelasticity of polydisperse polymers based on that of monodisperse systems were evaluated with simulation results.
\end{abstract}
\maketitle
\end{CJK*}

\newpage
\section{Introduction}\label{Sec_intro}
	The rheology of polymeric materials depends on their molecular structure. This dependence has been exploited in practical applications for improving melt processability and in characterizing their molecular weight distributions (MWD) \cite{dealy2018structure}. The reliability of these applications requires robust molecular theories capable of relating the rheological properties and chain relaxation dynamics to the underlying molecular make-up.

	For melts of short unentangled chains, their dynamics and viscoelasticity are well described by the Rouse model\cite{rouse1953theory, Bird_Curtis_1987, doi1988theory}. When the chains are sufficiently long and topological constraints between the chains are significant -- i.e. entangled chains, the tube model and reptation concept pioneered by \citet{edwards1967statistical} and \citet{de1971reptation} and further refined by \citet{doi1988theory} have been used with considerable success. Quantitative discrepancies between experimental results and the initial tube model have been attributed to the presence of additional relaxation mechanisms other than reptation\cite{VanRuymbeke2002}. Additional relaxation mechanisms such as contour length fluctuation (CLF)\cite{doi1988theory}, which describes the retraction and expansion of the contour length, and constraint release (CR)\cite{doi1987dynamics,des1992relaxation}, which accounts for the loss of entanglements due to the relaxation of other chains, have improved the quantitative prediction.

	These theories have largely been predicated on monodisperse samples. Industrial polymers are polydisperse and extension of these theories to polydisperse samples requires the understanding of the interplay between the relaxation dynamics of chains with different lengths. Semi-empirical mixing rules, which weigh the contributions of different chain lengths present in the polydisperse samples, are often used. The double reptation model tries to capture the effects of polydispersity naturally. Implicit in applying the double reptation model to polydisperse polymers is the assumption that the relaxation dynamics of each constituent chain length is not altered by the presence of other chains. Various stuides have shown that this is not the case \cite{Barsky2000,Kopf1997,Kalathi2014,shaffer1995effects}. Experimentally, it is difficult to generate, e.g., strictly bidisperse samples to study the mutual interaction between chains of different but well-defined molecular weights.
This problem naturally calls for
\RevisedText{molecular
simulation}
in which the constituent chain lengths can be precisely controlled.

\RevisedText{The charm of polymer physics problems lies in the fact that detailed chemical structures are often secondary to generic features of different polymers such as chain length and topology.
For this reason, important behaviors can often be captured in highly-simplified lattice models in Monte Carlo (MC) simulation.
\citet{baschnagel1998statics} used a bond-fluctuation model (BFM) to study bidisperse melts of short unentangled chains.
It was found that mixing chains of different lengths does not affect their statics (chain conformation statistics), but dynamics are noticeably shifted with the shorter chains becoming accelerated and longer chains slowed down.
This conclusion was corroborated by \citet{Lin2007} where a more delicate lattice model was used.
Unlike the BFM, which uses a cubic lattice, the newer model builds on a diamond lattice and the potential energy considers torsion angle variations according to the rotational isomeric state (RIS) model~\citep{rubinstein2003polymer}.
This setup allows for the modeling of chemically specific polymers.
The study investigated the dynamics of polyethylene chains in bidisperse melts at high temperature (\SI{453}{K}) where the longer species is well within the entangled regime.} 

\RevisedText{Molecular dynamics (MD) is the preferred method when realistic dynamics must be captured. Off-lattice models are typically used in MD simulation, among}
	which the bead-spring chain model pioneered by Kremer and Grest\cite{Kremer1990} has been particularly instrumental. However, most contributions focused on monodisperse polymer melts\cite{Kremer1990, kroger1993rheology, padding2002time,Likhtman2007a} and only a very small number of studies investigated the effects of polydispersity \cite{Barsky2000, Kopf1997, cao2010time, picu2007coarse, peters2018effect, peters2020viscoelastic}. A bidisperse system provides the simplest case where interactions between different chain lengths can be investigated. \citet{Barsky2000} studied chain diffusion dynamics in bidisperse mixtures of Kremer-Grest (KG) model chains and observed that mixing with longer chains reduces the mobility of the shorter chains while mixing with shorter chains accelerates the motion of longer chains.
	The longest chain considered in that study had $N = 90$ ($N$ is the number of monomeric units or "beads"), which is at most only marginally entangled and cannot capture most entanglement effects. In addition, limited by the computer power at the time, the study only probed the dynamics for a relatively short time period.
	Bidisperse systems of longer chains were studied later by Picu and Rakshit\cite{picu2007coarse} using a higher-level model which maps 40 KG beads into a single coarse-grained bead. Topological constraints due to entanglement are modeled by forcing the middle beads to move along the backbone (i.e., reptation) and only allowing three-dimensional motion in end beads. By construction, the model can only simulate well-entangled chains. They also found that the presence of surrounding shorter chains speeds up the dynamics of the longer chains. Because the model does not consider constraint release, the effect, at least as far their results are concerned, can only be attributed to chain-end effects.
	\RevisedText{\citet{Baig2010} also reported that mixing with shorter chains speeds up the relaxation of longer chains (and vice versa).
	The study used a more realistic united-atom model for bidisperse \textit{cis}-1,4-polybutadiene melts, which retains both CLF and CR effects, and used the tube and segmental survival probability functions as indicators of the relaxation of entangled chains.
	Despite the changing dynamics, it also reported that static properties of the constraining tubes, including their primitive path conformations, are not affected by varying the surrounding chain length.}
	Polydisperse polymers with realistic, albeit very narrow, molecular weight distributions were simulated more recently using a coarse-grained model for polyethylene\cite{peters2018effect, peters2020viscoelastic}. It was found that keeping the weight-average molecular weight constant, chain dynamics accelerates with polydispersity. Mobility of the shortest chains increases much faster than the average mobility does with polydispersity, which leads to CR and faster terminal relaxation.

	\RevisedText{There has been a particular appetite for probe diffusion problems, which study the dynamics of a specific chain type, labeled as the probe species, in a matrix of (usually) another chain type.
	This is a special case of binary mixtures where the volume fraction of the probe type $\phi_\text{probe}$ is sufficiently low that probe-probe interactions are not important.
	Such a setup is designed for studying the effects of matrix chains on the probe.
	The example most relevant to our study is \citet{wang2008constraint} which used MD to investigate the dynamics of a long entangled chain species (primarily $N_\text{L}=350$) in a matrix of a shorter chain species spanning both unentangled and entangled regimes ($N_\text{S}=\;$\numrange{25}{160}).
	Both types were semiflexible and the long-chain volume fraction was kept at $\phi_\text{L}=0.15$ to minimize the entanglement between long chains.
	The study focused on the CR release effects, which leads to tube relaxation (described as the Rouse motion of the tube itself), with varying short-chain length $N_\text{S}$.
	More recently, \citet{shanbhag2020molecular} used a BFM, MD, and slip-spring model to simulate the diffusion of a probe chain in two types of matrices, one with the same chain type as the probe and the other with infinitely long chains, designed for studying the self-diffusion and tracer diffusion of the probe chain, respectively.
		}

	\RevisedText{The general conclusion that mutual interactions in a binary mixture result in the acceleration and deceleration of the slower and faster chain species, respectively, not only applies to mixtures with different chain lengths, but also to those of different chemical types, which, for example, was also reported by \citet{Kopf1997} where the two chain species differ in monomeric mass.
	Interestingly, varying chain topology can lead to more complex mixing behaviors.
	Using a BFM, \citet{Shanbhag2017} showed that the mobility of a ring probe polymer changes non-monotonically with increasing length of the linear matrix chains.}	

	In this work, we investigated the dynamics and rheology of bidisperse polymer melts
\RevisedText{using the KG bead-spring chain model.}
\RevisedText{MD simulation was performed for binary} mixtures between
\RevisedText{an entangled chain species ($N_\text{L}=350$) and a shorter, unentangled or marginally entangled, chain species ($N_\text{S}=\;$\numlist{25;50;100} --}
monodisperse melts of the first two are unentangled and the last one, as shown below, is marginally entangled%
\RevisedText{), as we were particularly interested in}
the interplay between chains of different dynamical regimes.
\RevisedText{The chain lengths under our investigations fall into the same range as studied in \citet{wang2008constraint}. However, we studied compositions where neither component can be viewed as the probe -- i.e., each chain interacts with other chains of both the same and the opposite type.}
Two concentration levels were considered
\RevisedText{for each $N_\text{L}$--$N_\text{S}$ combination}
-- one has the $N_\text{L}=350$ species as the majority ($\phi_\text{L}=0.7$) and the other as the minority ($\phi_\text{L}=0.3$) component.
\RevisedText{Compared with previous studies on bidisperse melts with the KG model~\citep{Barsky2000,wang2008constraint}, our MD simulations were also substantially longer to cover the entire relaxation spectrum of all chains involved as well as to directly compute the full stress relaxation profile.}

 We started with the mean-squared displacement as a direct measurement of individual chain dynamics (section \ref{subsec:MSD}). It was followed by a Rouse mode analysis which reveals departure (or the lack thereof) from the unentangled limit and onset of entanglement in different components of the mixture (section \ref{subsec:RMA}). Finally, we examined the stress relaxation dynamics of the entire melt and evaluated mixing rules for predicting the relaxation modulus of the bidisperse system given that of the monodisperse melt of each component (section \ref{subsec:rheology}).
	
\section{Models and Numerical Methods} 
	The KG bead-spring model \cite{Kremer1990} was used. Each chain consists of $N$ beads bonded by finitely extensible non-linear elastic (FENE) springs~\citep{Bird_Curtis_1987}.
	The potential between bonding beads is
	\begin{align}
	U_\text{FENE}(r) &= -\frac{1}{2}K{R_0}^2\text{ln}\left[1-{\left(\frac{r}{R_0}\right)}^2\right] \notag \\
		&+4\epsilon\left[{\left(\frac{\sigma}{r}\right)}^{12}-{\left(\frac{\sigma}{r}\right)}^6+\frac{1}{4}\right]
		\label{eq:fene}
	\end{align}
\RevisedText{where} $r$ is the distance between the beads and $\sigma$ and $\epsilon$ are the standard Lennard-Jones (LJ) length and energy parameters. The first term of the equation represents an attractive potential which models FENE springs between nearest neighbors along the chain with a maximum bond length $R_0 = 1.5\sigma$, while the second term models the
excluded-volume repulsion 
	between beads and the term 
	is only included at $r \leq 2^{1/6}\sigma$. The spring
	constant 
	$K = 30\sigma/\epsilon$ is chosen to allow a reasonable integration time step while preventing chains from crossing each other\cite{Kremer1990}.
	\RevisedText{Note that compared with the semiflexible chains studied in \citet{wang2008constraint}, our model is fully flexible with no angle potential.
	This difference must be kept in mind for any comparison we make with that study below.}
	
	The interaction between non-bonding beads
	is 
	modeled by the standard LJ potential
	\begin{equation}
		U_\text{LJ}(r)=	4\epsilon\left[{\left(\frac{\sigma}{r}\right)}^{12}-{\left(\frac{\sigma}{r}\right)}^{6}\right].
		\label{eq:lj}
	\end{equation}
The potential is truncated at $r=2.5\sigma$ and shifted by a constant to ensure continuity at the cutoff.
	\RevisedText{Note that the original model by \citet{Kremer1990} used a shorter cutoff of $r = 2^{1/6}\sigma$, making the interaction between non-bonding beads purely repulsive.
	This practice is still widely seen in the polymer dynamics literature, although full LJ potential including the attraction well is used more often recently~\citep{grest2016communication, kalathi2014rouse, makke2011predictors, zhang2017effects}.
	Practically, comparison between the two approaches found no significant difference in both chain statics and dynamics when temperature is sufficiently high~\citep{kalathi2014rouse, grest2016communication} -- such as $T=1.0\epsilon/k_\text{B}$ used in this study ($k_\text{B}$ is the Boltzmann constant).}
	Hereinafter, all results will be reported in reduced LJ units in which length, energy, time, and temperature are scaled by $\sigma$, $\epsilon$,
	$\tau = \sqrt{m\sigma^2/\epsilon}$ ,
	and $\epsilon/k_\text{B}$, respectively.
	\RevisedText{For example, the non-dimensional LJ energy and length parameters in \cref{eq:fene} and \cref{eq:lj} are both unity.}
	A constant time step of 0.01 (in LJ time units or TUs) is used for all simulation.
	
	Each monodisperse system contains \RevisedText{\num{50000}} beads and each bidisperse system contains \RevisedText{\num{56000}} beads.
	\RevisedText{The difference in size here is very small and, for selected monodisperse cases, we have tested the larger size of \num{56000} and found no noticeable size dependence.
	For both mono- and bidisperse systems,}
	the beads were placed in a cubic box with periodic boundary conditions at a constant bead density of 0.85. The bidisperse systems mix a long-chain species ($N_\text{L}=350$) with shorter chains of various lengths ($N_\text{S}=25,50$, or $100$) at two levels of mass (or volume -- the constituent beads are identical) fraction: $\phi_\text{L}=0.7$ and $0.3$. The longer chain is moderately entangled --for reference, the entanglement strand length $N_\text{e}$ for the KG model is in the range of 30 to 80 depending on the method of
	determination
	\cite{xi2019molecular}. Detailed compositions are listed in Table \ref{Model A}.
	
	All the simulations were carried out using the Large-scale Atomic/Molecular Massively Parallel Simulator (LAMMPS) package \cite{plimpton1993fast}.
	The initial configuration was generated by randomly placing the specified number and types of chains in the simulation cell. Generation of individual chains follows a procedure that is analogous to a self-avoiding walk in a continuum space, which prevents backfolding of successive bonds but still leaves a large number of bead overlaps.
	A dissipative particle dynamics (DPD) push-off method\cite{zhang2017effects}, originally proposed by \citet{sliozberg2012fast}, was then used to obtain an equilibrated structure for production runs.
	During the DPD run, interaction between non-bonding beads was replaced by	
	a \RevisedText{repulsive potential in} the form of
	\begin{align}
		U_\text{DPD}(r) = 
		\begin{dcases}
			\frac{A_\text{DPD}}{2}r_c\left(1-\frac{r}{r_c}\right)
				& r < r_c\\
			0 & r\geq r_c
		\end{dcases}.
	\end{align} 
	\RevisedText{Compared with the LJ potential (\cref{eq:lj}), which has nearly hard-sphere repulsion at short range, the DPD potential is much softer. It allows for easier passing of chains and thus fast relaxation during the initial equilibration steps.}

	DPD simulation was run at $T = 1.0$ using
	a cut-off distance $r_c = 1.0$. The potential was initially low with $A_\text{DPD}=25$. At the beginning, restriction was imposed on the maximum distance each bead can move within one time step which gradually increases from 0.001 to 0.1 over $\SI{15}{TUs}$. The restriction was then removed and the simulation was run for another $\SI{100}{TUs}$. This was subsequently followed by a gradual ramp of $A_\text{DPD}$ to 1000 over $\SI{5.5}{TUs}$. The DPD potential was then replaced with the standard LJ potential and MD in an NVT ensemble was performed for additional $\SI{500}{TUs}$ during which a random velocity distribution was reassigned to all beads every $\SI{0.5}{TUs}$.
	
	\begin{table}
		\centering
		\caption{Compositions of bidisperse systems simulated:
		$N$, $\phi$, and $n_\text{c}$ denote the chain length, mass/volume fraction, and total number of chains of a given species, respectively; subscripts ``L'' and ``S'' denote the longer and shorter component in the mixture, respectively.}
		\label{Model A}
		\begin{tabular}{ccc|ccc|c}
			\hline
			$N_\text{L}$ \qquad & $\phi_\text{L}$ \quad &
			$n_\text{c, L}$ 
			\quad & $N_\text{S}$ \quad & $\phi_\text{S}$ \quad &
			$n_\text{c, S}$ 
			\quad& Total beads \\
			\hline
			\hline
			350 & 0.7  & 112 & 25 & 0.3  & 672 & 56000 \\
			350 & 0.7  & 112 & 50 & 0.3  & 336 & 56000 \\
			350 & 0.7  & 112 & 100 & 0.3  & 168 & 56000\\
			350 & 0.3  & 48  & 25  & 0.7  & 1568 & 56000\\
			350 & 0.3  & 48  & 50  & 0.7  & 784  & 56000\\
			350 & 0.3  & 48  & 100 & 0.7  & 392  & 56000\\
			\hline
		\end{tabular}
	\end{table}

\begin{figure}
		\centering
		\includegraphics[width=3.5in, trim=0 0 0 0, clip]{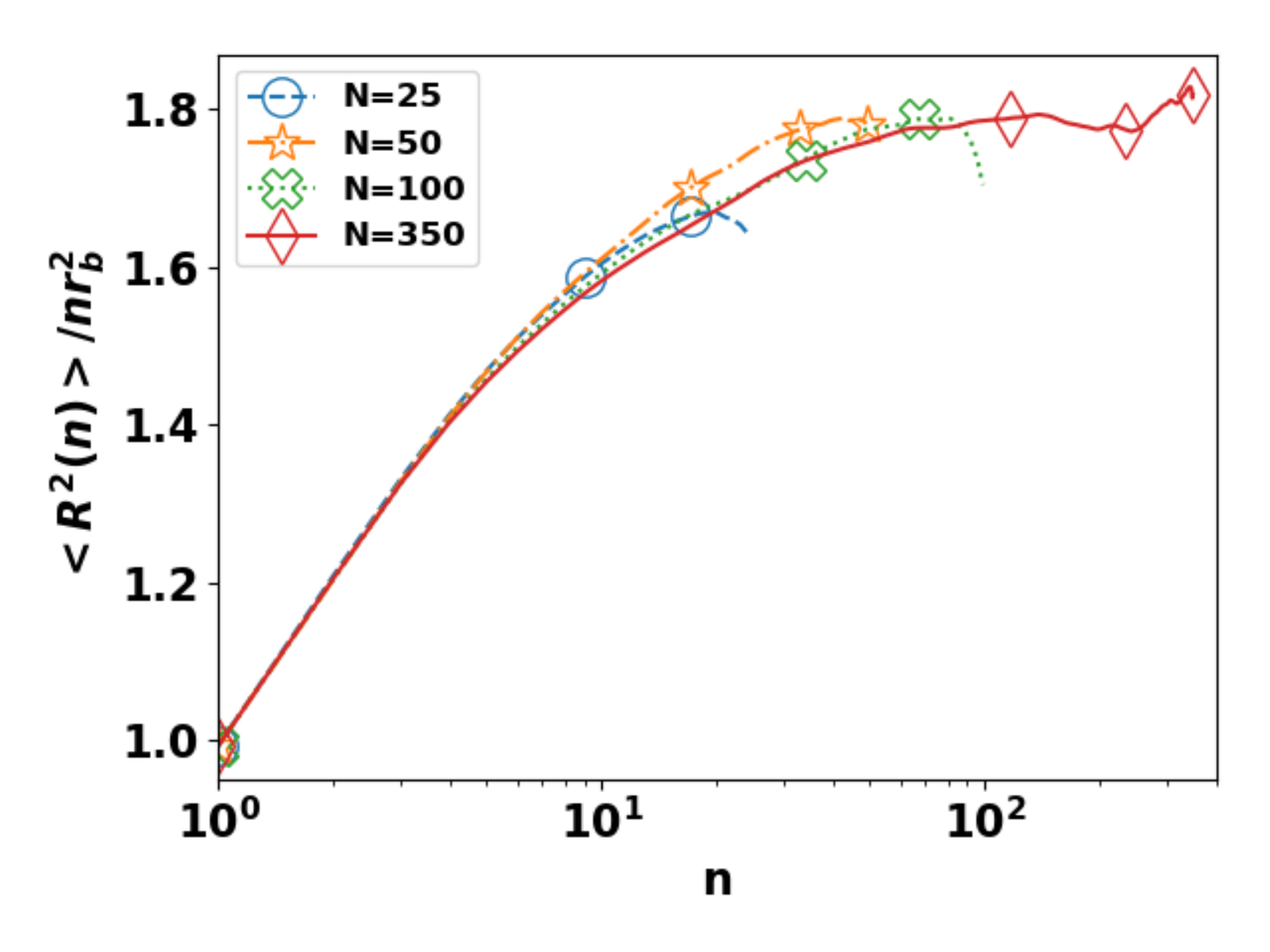}
		\caption{Characteristic ratio calculated from the mean square internal displacement (MSID) for equilibrated monodisperse systems of different chain lengths.}
		\label{MSID_plot}
	\end{figure}

	Equilibration quality was
	examined through the mean square internal displacement (MSID)
	\begin{gather}
		\left\langle R^2(n)\right\rangle\equiv\left\langle|\vec r_j-\vec r_i|^2\right\rangle
	\end{gather}
	which measures the square distance between the $i$-th and $j$-th monomeric unit of the same chain, averaged over all $ij$-pairs with the same index separation $n\equiv|j-i|$. \citet{auhl2003equilibration} showed that, compared with the radial distribution function, end-to-end distance and radius of gyration, MSID better captures chain deformation at intermediate scales which does not fully relax until the whole chain is equilibrated. Figure \ref{MSID_plot} plots $\langle R^2(n)\rangle/(nr_b^2)$ ($r_b=0.97$ is the equilibrium bond/spring length), which is the characteristic ratio of the KG chain, versus $n$ for our equilibrated monodisperse systems. All curves increase monotonically at the beginning. For the longer chains ($N \geq 50$), the characteristic ratio converges to a constant value at the large $n$ limit -- i.e., $C_\infty$. The obtained $C_\infty = 1.75$ in our simulation is in excellent agreement with previous studies (e.g., 1.74 in \citet{Kremer1990} and 1.75 in \citet{auhl2003equilibration}).
	
	For each cell composition, three random initial configurations were independently generated and each underwent the above equilibration procedure before its production run.
	\RevisedText{Productions runs were performed in an NVT ensemble with $T=1.0$ using Nose-Hoover chains for thermostating.}
	The production run
	of each configuration
	lasted for a total of
	$\SI{5e5}{TUs}$ for monodisperse $N=25$ and $N=50$ cases,
	$\SI{1e6}{TUs}$ for the monodisperse $N=100$ case, and
	$\SI{3e6}{TUs}$ for all other cases (any system, monodisperse or bidisperse, containing $N=350$ chains).
	Results reported in this study were averages over these three trajectories from independent initial configurations unless specified otherwise. Error bars\RevisedText{, when provided,} report the standard error between the independent runs. 
	
\section{Results and Discussion}
\subsection{Mean Square Displacement (MSD)}\label{subsec:MSD}
	We studied the dynamics of each component in the mixture using the monomer MSD $g_1(t)$ which describes the motion of individual monomers
	\begin{equation}
		g_1(t) \equiv \frac{1}{n_c}\sum_{i=1}^{n_c} \left(\frac{1}{N_j}\sum_{j=1}^{N_i}[\vec{r}_{ij}(t)-\vec{r}_{ij}(0)]^2\right)
		\label{g1t}
	\end{equation}
	and the MSD of the center of mass $g_3(t)$ which describes the overall motion of the center of mass of individual chains
	\begin{equation}
		g_3(t)\equiv\frac{1}{n_\text{c}}\sum_{i=1}^{n_\text{c}}[\vec{r}_{i,c.m.}(t) - \vec{r}_{i,c.m.}(0) ]^2
	\end{equation}
	where
	\begin{equation}
		\vec{r}_{i,\text{c.m.}}(t)\equiv\frac{1}{N_i}\sum_{j=1}^{N_i}\vec{r}_{i,j}(t),
	\end{equation}
	$n_c$ is the number of chains of each component, $N_i$ is the length of the $i$-th chain, $\vec{r}_{i,j}(t)$ is the position of the $j$-th monomer of the $i$-th chain, and $\vec{r}_{i,\text{c.m.}}(t)$ is the position of the center of mass of the $i$-th chain \cite{kremer1992simulations,Hsu2016}. An efficient method for MSD calculation based on fast Fourier transform (FFT), as detailed in \citet{press1992numerical} and \citet{calandrini2011nmoldyn}, was used. 

	Theoretically, the monomeric MSD of an unentangled chain described by the Rouse model follows the following scaling behavior in different time regimes~\cite{doi1988theory, Hsu2016}
	\begin{equation}	
		g_1(t) \sim 
		\begin{dcases}
			t^1 & t < \tau_0\\
			t^{1/2} & \tau_0 < t < \tau_\text{R}\\
			t^1 &  t > \tau_\text{R}
		\end{dcases},
		\qquad
		\label{g1t_prediction}
	\end{equation}
	where $\tau_0$ signifies the characteristic relaxation time of a single monomer and $\tau_\text{R}$ is the Rouse time -- the characteristic relaxation time of the whole chain (according to the Rouse model). For entangled systems, the \RevisedText{simple tube model} gives the following scalings
	\begin{gather} 
		g_1(t) \sim 
		\begin{dcases}
			t^1 & t < \tau_0\\
			t^{1/2}& \tau_0 < t < \tau_\text{e}\\
			t^{1/4} & \tau_\text{e} < t < \tau_\text{R}\\
			t^{1/2} & \tau_\text{R} < t < \tau_\text{d}\\
			t^{1} & t > \tau_\text{d}
		\end{dcases}
		\label{eq:g1_prediction}
	\end{gather}
	and
	\begin{gather}
		g_3 (t) \sim 
		\begin{cases}
			t^1 & t < \tau_\text{e}\\
			t^{1/2} & \tau_\text{e} < t < \tau_\text{R}\\
			t^{1} & t > \tau_\text{R}\\
		\end{cases}
		\label{eq:g3_prediction}
	\end{gather}
	where $\tau_\text{e}$ is the relaxation time of an entanglement strand and $\tau_\text{d}$ is the disentanglement time
	-- the longest relaxation time of an entangled chain.

\begin{figure}
		\centering
		\includegraphics[width=3.5in, trim=0 0 0 0, clip]{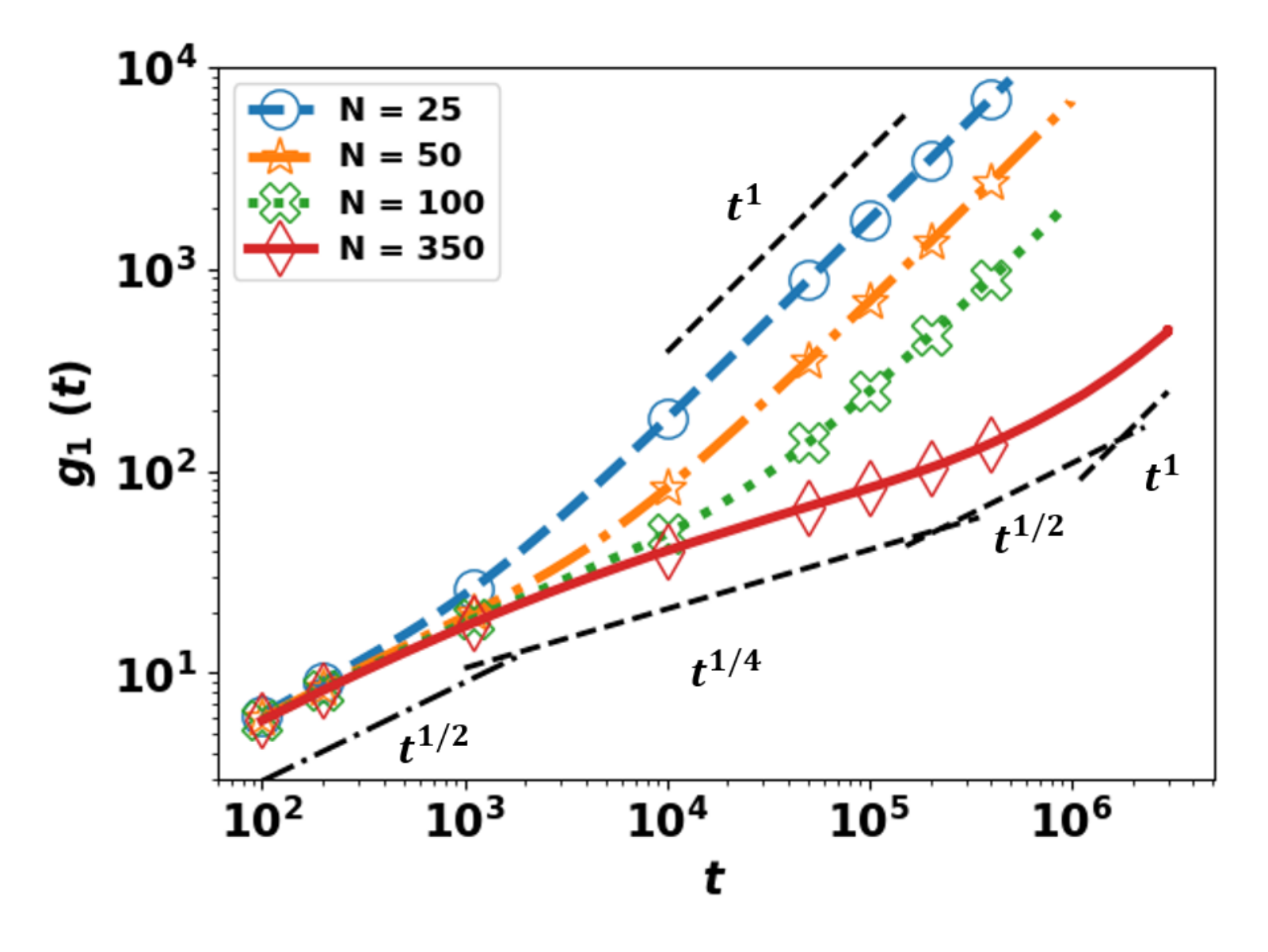}
		\caption{Mean square displacement of internal monomers $g_1(t)$ of monodisperse systems.}
		\label{fig:g1t_mono}
	\end{figure}

	The monomer MSD of our monodisperse systems is presented in fig. \ref{fig:g1t_mono}. Our calculation of $g_1(t)$ did not average over all monomers as indicated in eq. \ref{g1t}. To minimize chain-end effects, only the middle monomer of each chain was included. At early times, all chains should start out with $g_1(t)$ scaling with $t^1$. This regime is not captured in fig. \ref{fig:g1t_mono} because our sampling frequency was not high enough to capture the dynamics at such a small time scale. The next expected regime, for both unentangled and entangled chains, has a $t^{1/2}$ scaling which is indeed observed in all chain lengths studied. Dynamics of different chain lengths diverge thereafter. The $N=25$ and $50$ cases directly enter the diffusive regime ($t^1$), which is consistent with the Rouse model prediction (eq. \ref{g1t_prediction}). Note that for $N = 50$, complete alignment with the $t^1$ scaling starts at $t\sim\mathcal{O}(10^4)$ which agrees with previous studies of monodisperse polymer dynamics\cite{Kremer1990}. In comparison, the earlier bidisperse polymer study by \citet{Barsky2000} only covered time scales up to $t\approx 4000$. Deviation from this pure Rouse dynamics is seen in longer chains. For $N = 100$, we observe a clear slow down in the $t\sim\mathcal{O}(10^3)$ to $\mathcal{O}(10^4)$ regime, but it falls short of completely dropping to a $t^{1/4}$ scaling, which reflects weak entanglement and an insufficient separation between the entanglement strand length and the chain length (i.e., insufficient separation between $\tau_\text{e}$ and $\tau_\text{R}$). A pronounced $\tau_\text{e}<t<\tau_\text{R}$ regime is found in the longest chain with $N = 350$.
	A least-square regression analysis of the MSD data from $t\approx 3\times10^3$ to $2\times 10^5$ \RevisedText{gives} a $t^{0.28\pm 0.02}$ scaling, which is close to, \RevisedText{but still slightly higher than}, the theoretical $t^{1/4}$ prediction. 

	\RevisedText{For comparison, the $g_1(t)$ profile for monodisperse $N=350$ semiflexible chains reported in \citet{wang2008constraint} was also visibly steeper than the $t^{1/4}$ scaling line. 
	The increased slope could be attributed to CR and CLF, both of which were not considered in \cref{eq:g1_prediction}.
	CR leads to the relaxation of the constraining tubes and the mobility of chain segments contains contributions from both chain reptation within tubes and the tube Rouse motion~\citep{viovy1991constraint}.
	Meanwhile, even the longest $N=350$ chains studied here are still not significantly longer than the entanglement threshold. Therefore, CLF is felt over substantial portions of the chains (if not the entire chains), instead of just the chain ends (as in the case of well-entangled chains).
	Theories are typically constructed for much longer chains where the number of entanglement strands per chain $Z\equiv N/N_\text{e}\gg\mathcal{O}(1)$ (see $N_\text{e}$ estimated below in \cref{eq:Ne}).
	Finally, we also note that slope estimation from regression is always subject to statistical error and can also be sensitive to the range of data points used.
	The increase seen in our $g_1(t)$ profile slope for $N = 350$ is not large compared with uncertainty.}

	By fitting different segments of MSD data to the scaling laws with their corresponding theoretical exponents (\cref{eq:g1_prediction}), we can identify the time scales for different dynamical regimes based on the intersections between the fitted lines.
	\RevisedText{To obtain $\tau_0$, we ran an additional short \SI{100}{TU} simulation with data stored at higher frequency for the $N=350$ case (to obtain the $g_1(t)$ profile for smaller time scales than those shown in \cref{fig:g1t_mono}).}
	The results are summarized in \cref{tab:timescales}.
	This approach, however, can be sensitive to the specific regression procedure and its uncertainty. From the obtained time scales , the entanglement strand length
	\begin{gather}
		N_\text{e}=\left(\frac{\tau_\text{e}}{\tau_0}\right)^{1/2}\approx 33
		\label{eq:Ne}
	\end{gather}
	can be estimated, which is consistent with literature values based on MSD\cite{Kremer1990}. 

	\begin{table}
		\centering
		\caption{Time scales extracted from the pure $N=350$ melt MSD curve.}
		\label{tab:timescales}
		\begin{tabular}{cccc}
			\hline
			\hline
			$\tau_0$	&$\tau_\text{e}$
				&$\tau_\text{R}$	&$\tau_\text{d}$	\\
			\hline
			$3.20$	&$\num{3.43e3}$	&$\num{1.66e5}$	&$\num{1.74e6}$	\\
			\hline
			\hline
		\end{tabular}
	\end{table}
	
\begin{figure}
		\centering
		\begin{subfigure}[b]{3.5in}
			\centering
			\includegraphics[width=\textwidth]{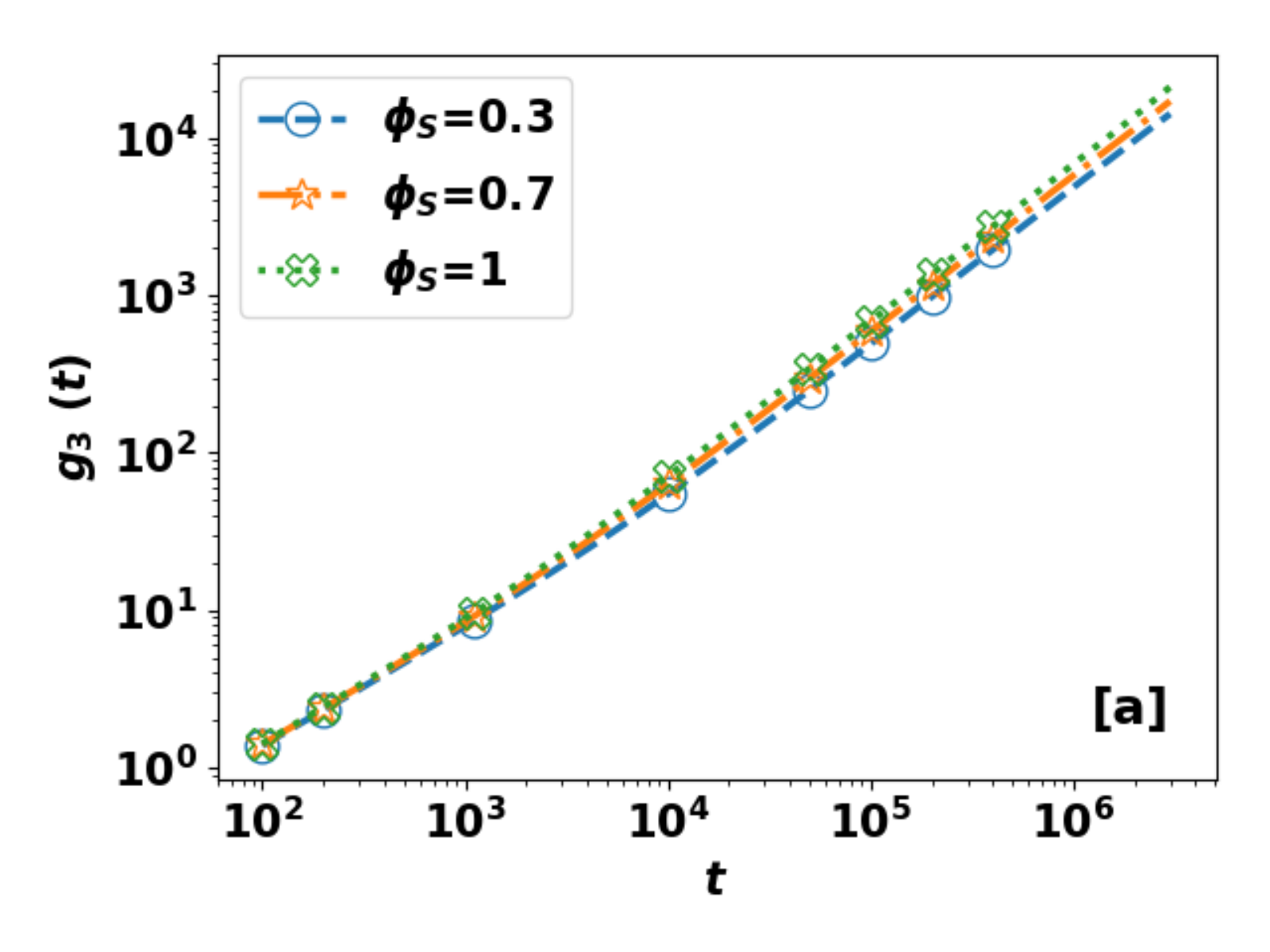}
			\phantomsubcaption
			\label{fig: g3t_50-350_pure_min_max}
		\end{subfigure}
		\begin{subfigure}[b]{3.5in}
			\centering
			\includegraphics[width=\textwidth]{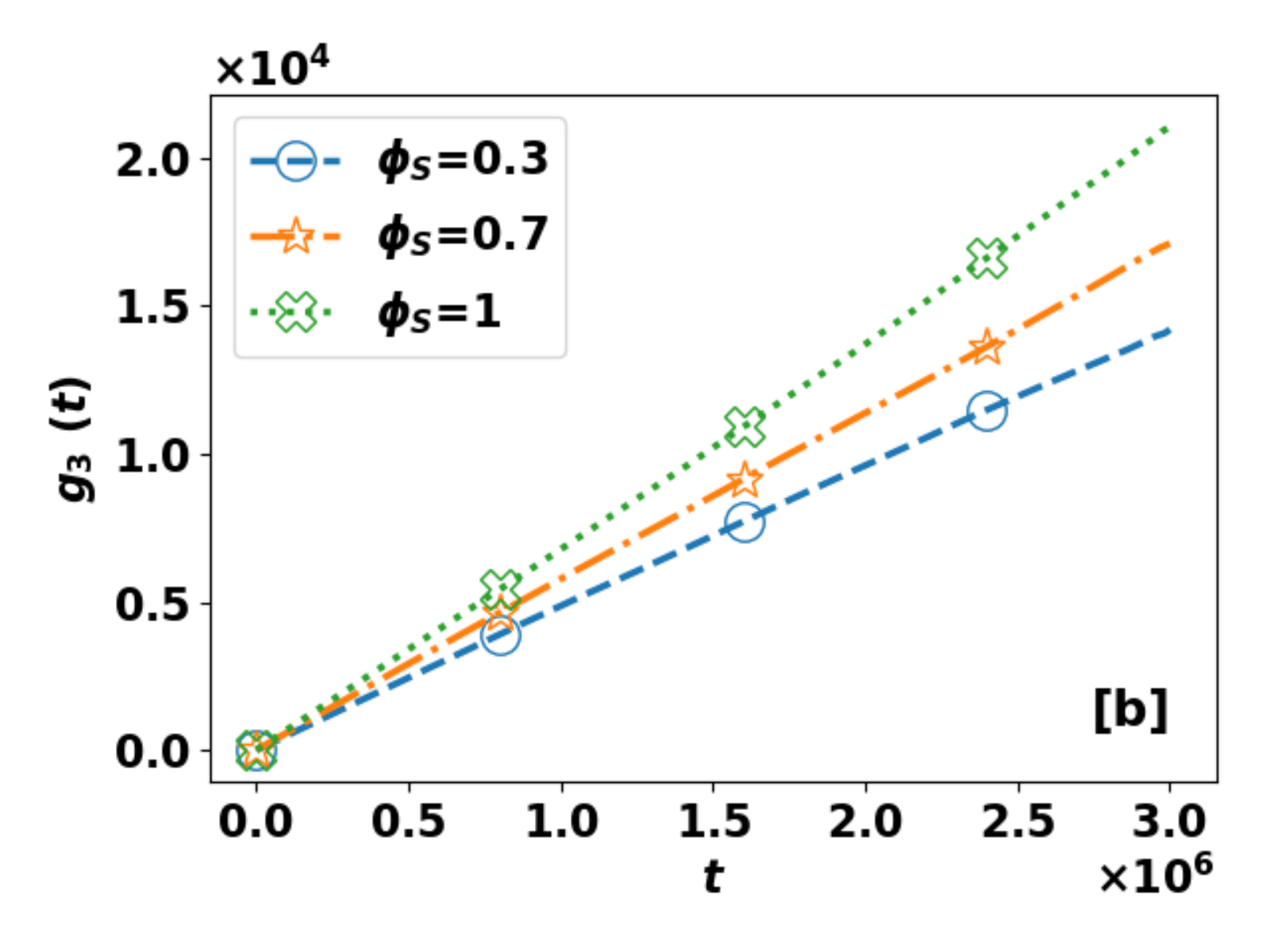}
			\phantomsubcaption 
			\label{fig:g3t_50-350_pure_min_max_no_log}
		\end{subfigure}
		\caption{Mean square displacement of the center of mass $g_3(t)$ for $N = 50$ chains in its pure melt and in bidisperse mixtures with $N_\text{L}=350$ chains as the minority and majority component:
		(a) log-log coordinates and (b) linear coordinates.}
	\end{figure}

	We next shift our focus to bidisperse systems, starting with the effects of a longer chain component ($N_\text{L}=350$) on the mobility of the shorter chains. \Cref{fig: g3t_50-350_pure_min_max} shows the center-of-mass MSD, $g_3(t)$, of pure $N = 50$ chains and the same $N_\text{S} = 50$ chains when mixed with a longer $N_\text{L} = 350$ component in log-log coordinates. At short time, MSD curves from mono- and bi-disperse systems appear indistinguishable. However, the pure $N = 50$ case ($\phi_\text{S}=1$) is the first to transition from the $t^{1/2}$ to $t^1$ scaling and shows higher chain mobility afterwards. With increasing concentration of the longer chain (lowering $\phi_\text{S}$), mobility of the $N = 50$ chains decreases. The difference appears small in \cref{fig: g3t_50-350_pure_min_max}, but when put in linear scales (\cref{fig:g3t_50-350_pure_min_max_no_log}), it is clear that the diffusion rate of the $N = 50$ chains decreases.
	The same observation,
	that the long-chain component impedes the motion of the shorter chains, is also made when other shorter chains ($N_\text{S} = 25$ and $100$) are mixed with $N_\text{L} = 350$ chains.
	\RevisedText{It also agrees with the general observation made in a number of previous studies~\citep{baschnagel1998statics,Barsky2000,Picu2007a,Baig2010}.}

\begin{figure}
		\centering
		\begin{subfigure}[b]{3.5in}
			\centering
			\includegraphics[width=\textwidth]{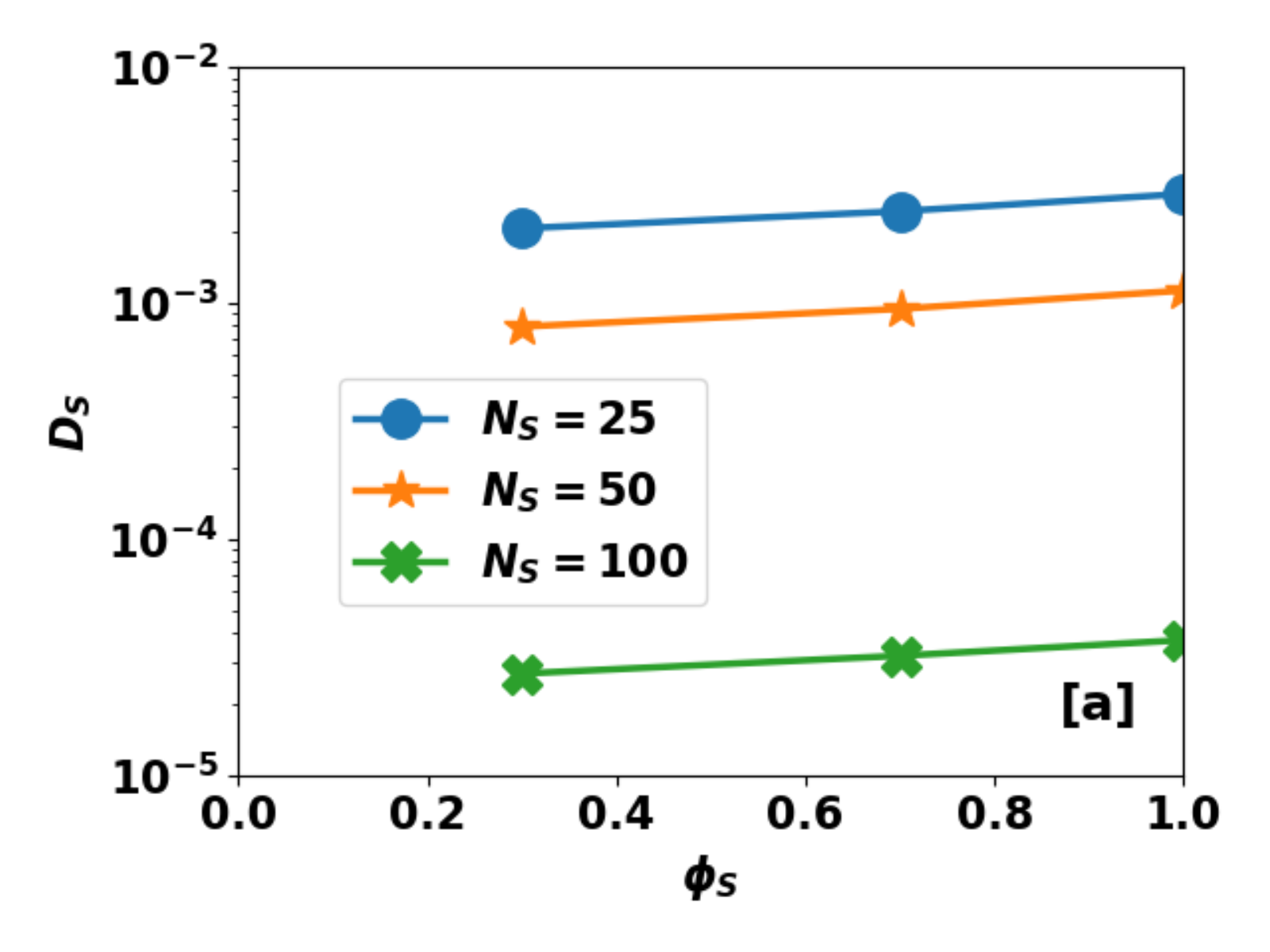}
			\phantomsubcaption
			\label{fig:diff_short_phi}	
		\end{subfigure}
		\begin{subfigure}[b]{3.5in}
			\centering
			\includegraphics[width=\textwidth]{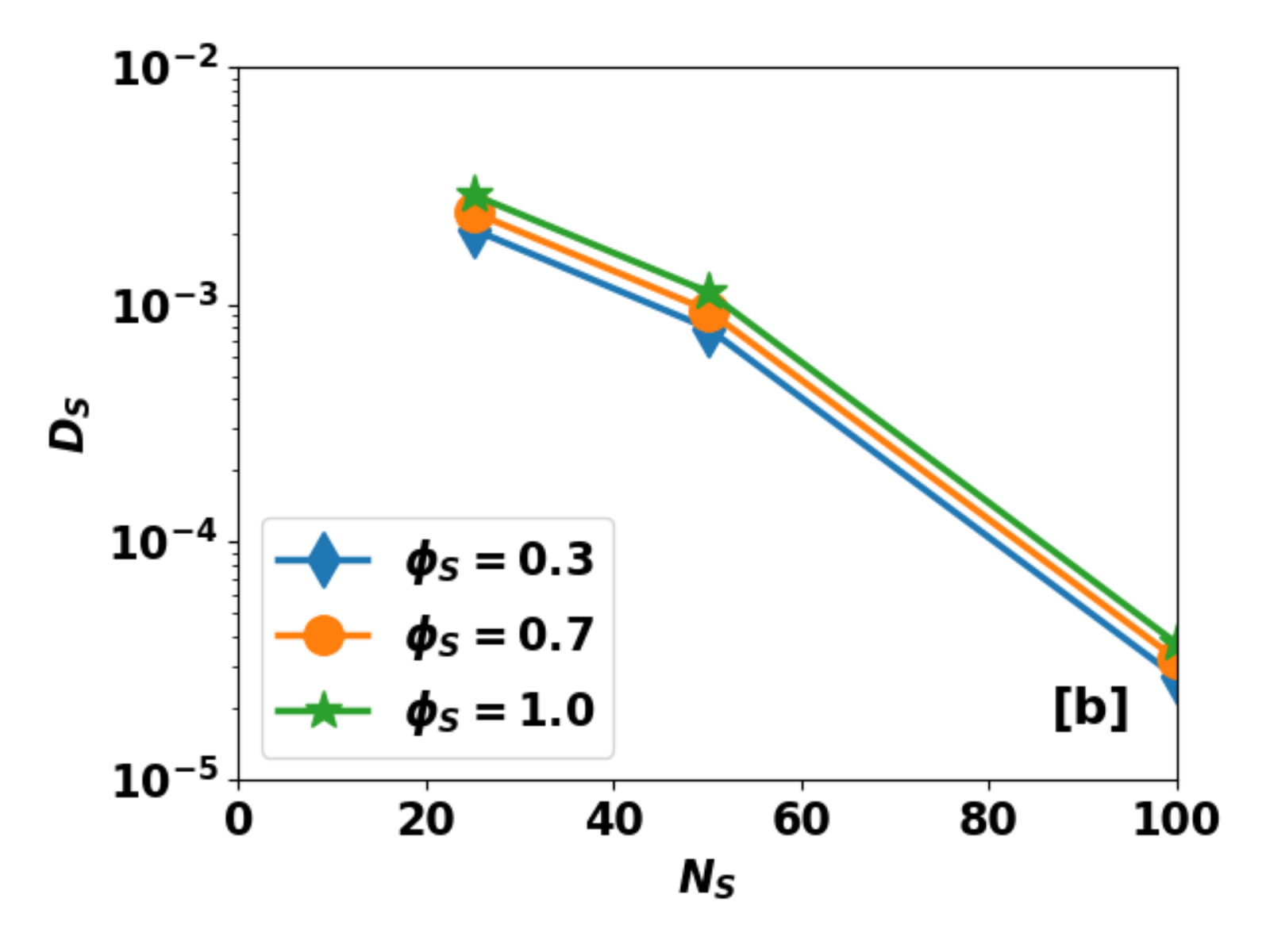}
			\phantomsubcaption 
			\label{fig:diff_short_N}
		\end{subfigure}
		\caption{Diffusion coefficient of the short-chain component in bidisperse mixtures with $N_\text{L}=350$ chains (compared with that of pure short chain), as functions of (a) short chain mass/volume fraction $\phi_\text{S}$ and (b) short chain length $N_\text{S}$}	
	\label{fig:diff_short_plot}
	\end{figure}

	We further quantify chain mobility in the long-time limit by calculating its self-diffusion coefficient $D$ using the Einstein relation
	\begin{equation}
		\lim_{t\to\infty} g_3(t)\sim 6Dt
	\end{equation}
	which allows the extraction of $D$ from the slope of the MSD curve.
	Figure \ref{fig:diff_short_plot} shows the diffusion coefficient of the shorter component in bidisperse mixtures as well as that of pure short-chain melts. As shown in figure \ref{fig:diff_short_phi}, the presence of the longer chains ($N_\text{L} = 350$) substantially reduces the diffusion rate of the shorter chains regardless of the length of the latter and the effect is stronger as the fraction of the longer component increases. The change in $D_\text{S}$ may not appear large in the logarithmic scale used in the figure, but for $\phi_\text{S} =0.7$ and $0.3$, $D_\text{S}$ drops by $\approx 15\%$ and $30\%$, respectively, for all three short chain length $N_\text{S}$ levels.

	Figure \ref{fig:diff_short_N} re-plots the same set of data using $N_\text{S}$ as the independent variable. It highlights that for the three composition levels (including the pure short-chain limit) studied, the \RevisedText{chain-length} dependence follows the same pattern
	\RevisedText{-- i.e., lines for different $\phi_\text{S}$ stay parallel to one another.
	The trend extends at least up to $N_\text{S}=100$ where weak entanglement has already kicked in, as reflected by the steeper slope between $N_\text{S}=50$ and $100$.}
	This means that, e.g., $D_\text{S}(N_\text{S}=50)/D_\text{S}(N_\text{S}=25)$ stays nearly the same for different $\phi_\text{S}$ levels (note, again, the logarithmic scale in $D_\text{S}$), at least in the range tested, suggesting that the increasing resistance brought by the long chain can be lumped into a monomeric friction factor that increases with \RevisedText{the} long-chain fraction \RevisedText{but remains independent of $N_\text{S}$ -- i.e.,} $\zeta(\phi_\text{L})$. This observation can be rationalized considering that the relaxation of the longer species is a much slower process and, within the relaxation time of the shorter component, the long chains can be approximated as an invariant matrix.

\begin{figure}
		\centering
		\includegraphics[width=3.5in, trim=0 0 0 0, clip]{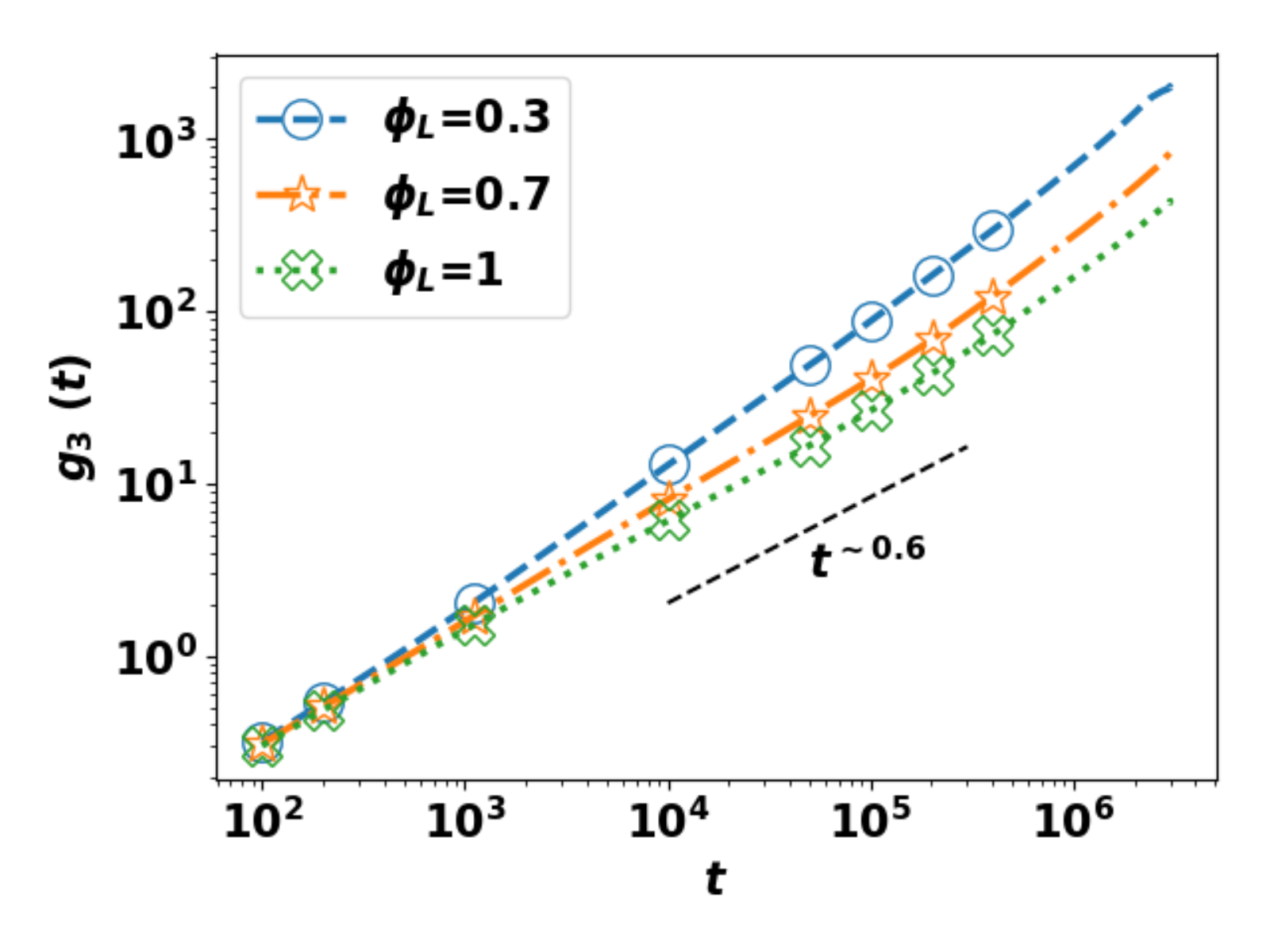}
		\caption{MSD of the center of mass $g_3(t)$ of $N_\text{L} = 350$ chains in bidisperse mixtures with $N_\text{S} = 25$ chains (both as the minority and as the majority component). MSD of pure $N = 350$ chains is also included for comparison.}
		\label{fig:g3t_350(25)_pure_min_max}
	\end{figure}

	We turn now to the effects of the shorter component on the long chains. Figure~\ref{fig:g3t_350(25)_pure_min_max} shows the center-of-mass MSD of $N_\text{L} = 350$ chains in bidisperse mixtures with the $N_\text{S} = 25$ species as the diluent. It is observed that the presence of the shorter component speeds up the relaxation of the longer chains and the effect increases with the short-chain fraction $\phi_\text{S}=1-\phi_\text{L}$. Since $N = 350\gg N_\text{e}$, entanglement effects are clearly shown in the $g_3(t)$ curve of the pure $N=350$ case -- a $t^{\sim 0.6}$ scaling regime is found around $t\sim O(10^4)$.
	The exponent is slightly higher than the theoretical prediction of $1/2$ in the $\tau_\text{e}<t<\tau_\text{R}$ regime (\cref{eq:g3_prediction}).
	\RevisedText{Again, we note that the $N=350$ semiflexible chains reported in \citet{wang2008constraint} also showed steeper $g_3(t)$ profile than the theoretical $t^{1/2}$ prediction.
	The discrepancy, as discussed above, may still be attributed to CLF and CR which are not considered in the theory.}
	Transition to the $t^1$ diffusive regime occurs at $t\approx 2 \times 10^{5}\sim\tau_\text{R}$. With increasing fraction of the short chains, the exponent (slope in log-log coordinates) of the same time range $t\sim O(10^4)$, greatly increases, reaching $0.82$ at $\phi_\text{L}=0.3$. This intermediate regime becomes less differentiable from the long-time diffusive limit, suggesting a weakening of entanglement effects by short chain addition.\par

\begin{figure}
		\centering
		\begin{subfigure}[b]{3.5in}
			\centering
			\includegraphics[width=\textwidth]{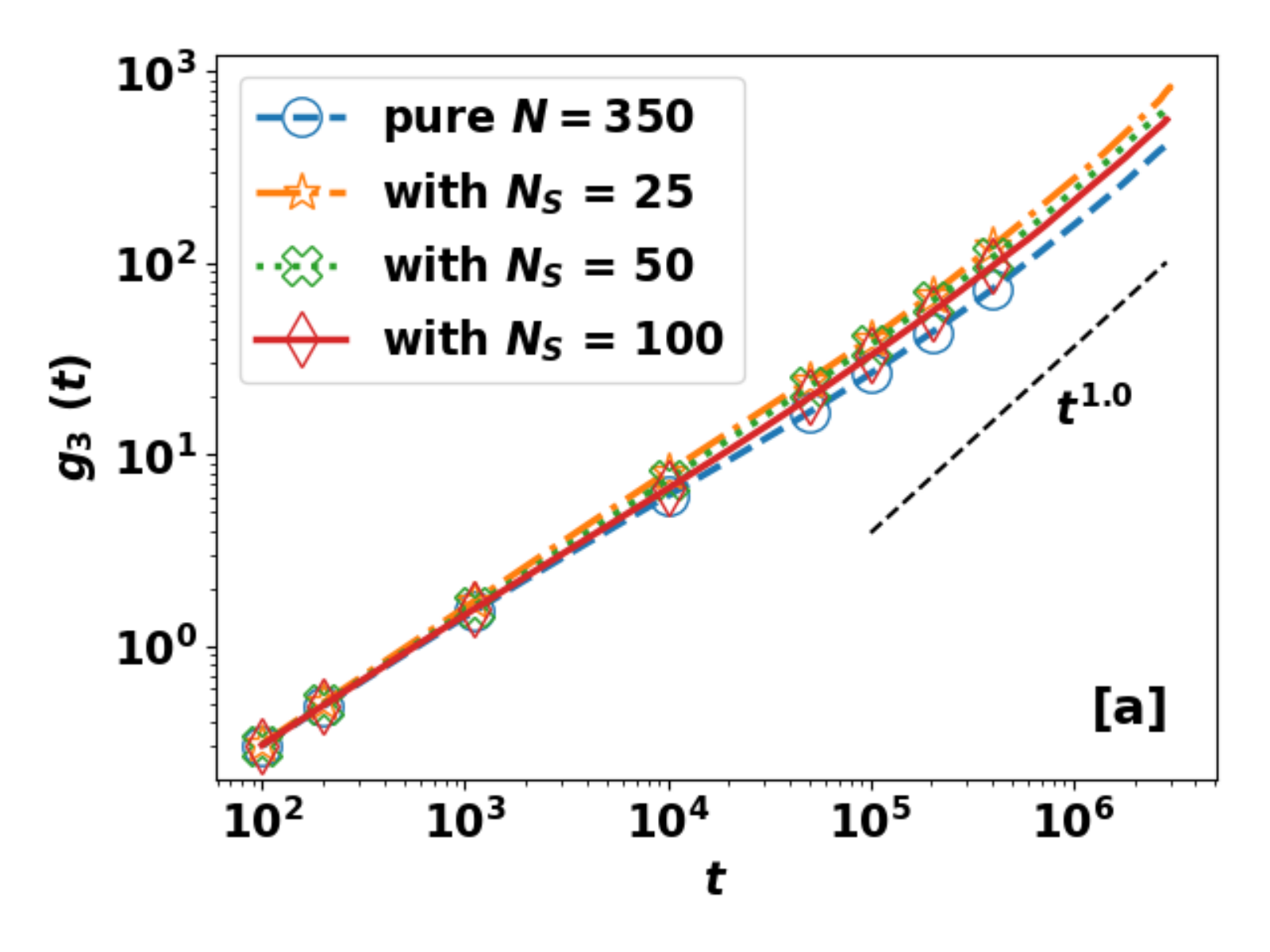}
			\phantomsubcaption
			\label{fig:350_majority_compared}
		\end{subfigure}
		\begin{subfigure}[b]{3.5in}
			\centering
			\includegraphics[width=\textwidth]{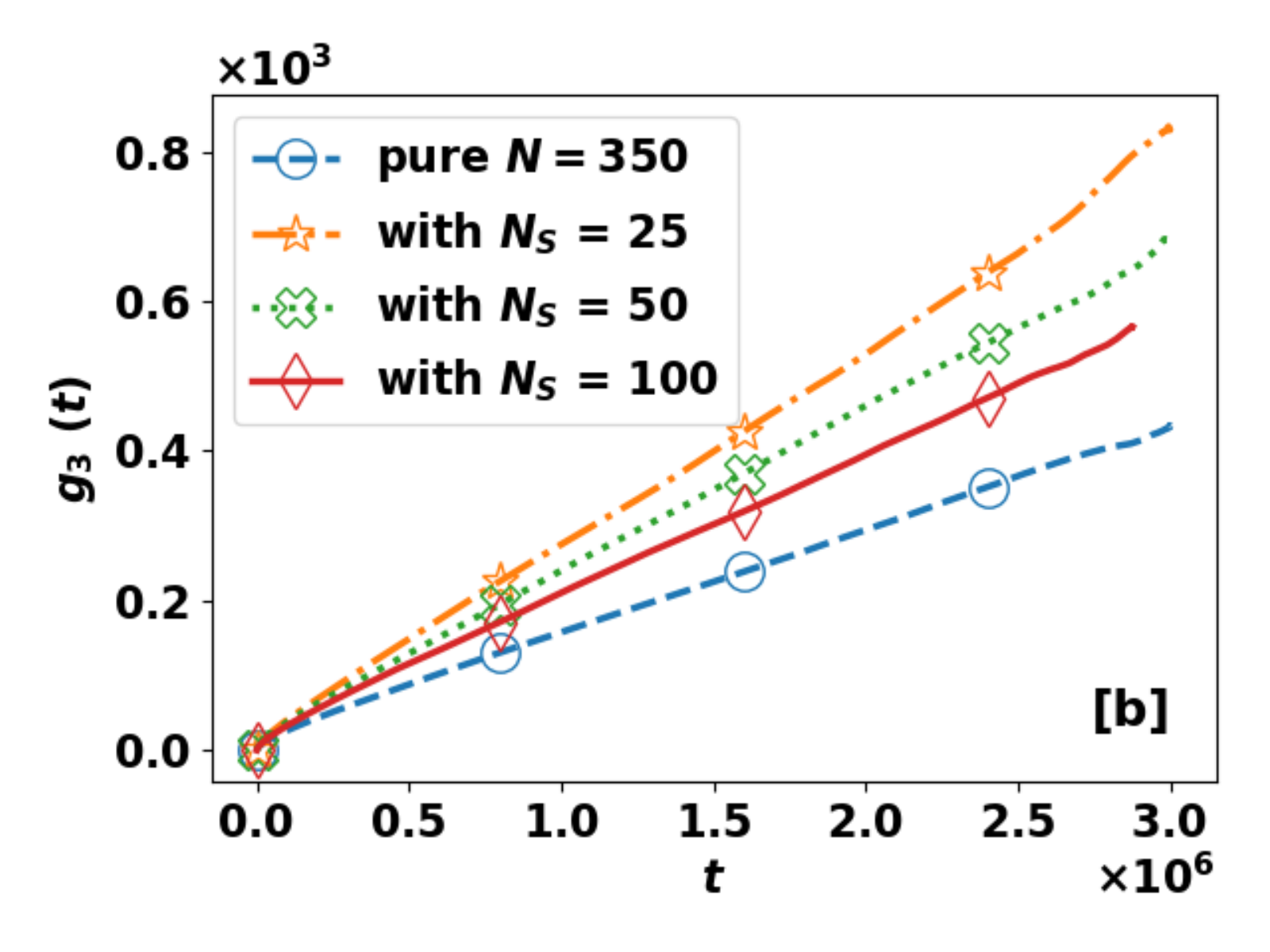}
			\phantomsubcaption
			\label{fig:350_majority_compared_no_log}
		\end{subfigure}
		\caption{MSD of the center of mass $g_3(t)$ of $N_\text{L} = 350$ chains in bidisperse mixtures as the majority component ($\phi_\text{L} = 0.7$), with $N_\text{S} = 25, 50$, and $100$ chains as the minority component: (a) log-log coordinates; (b) linear coordinates. The MSD of pure $N = 350$ chains is also included for comparison.}
	\end{figure}

\begin{figure}
		\centering
		\begin{subfigure}[b]{3.5in}
			\centering
			\includegraphics[width=\textwidth]{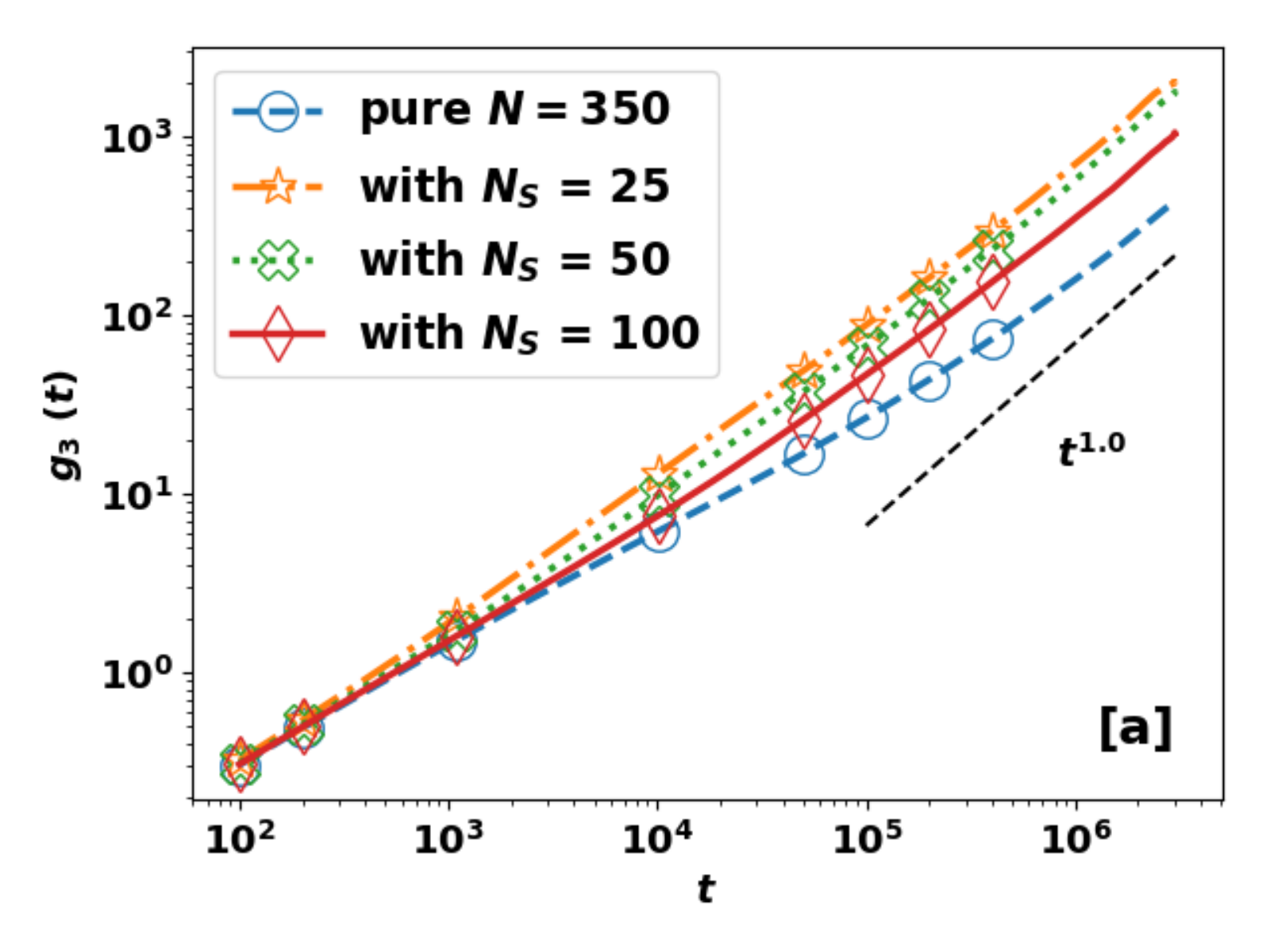}
			\phantomsubcaption
			\label{fig:350_minority_compared}
		\end{subfigure}
		\begin{subfigure}[b]{3.5in}
			\centering
			\includegraphics[width=\textwidth]{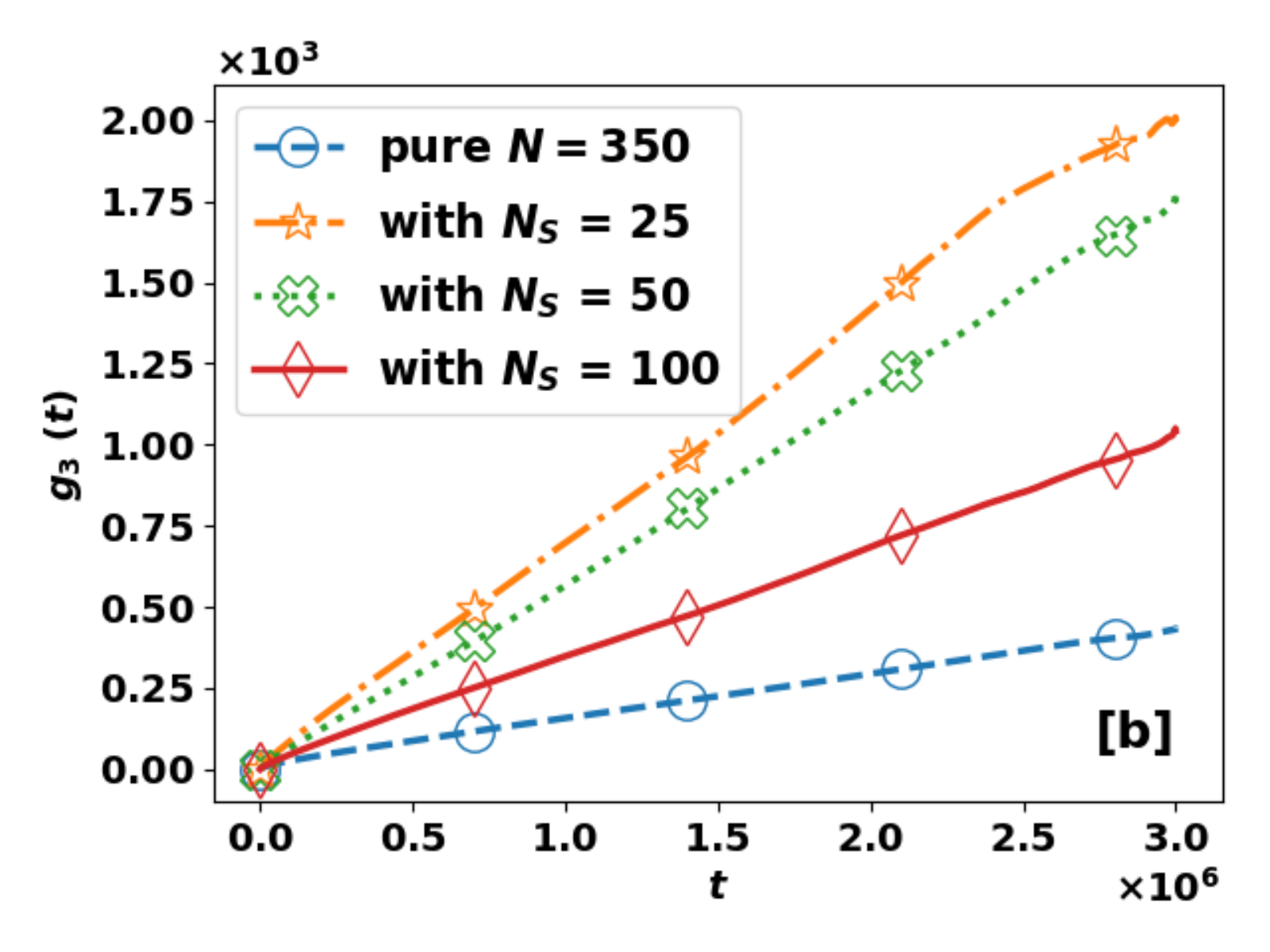}
			\phantomsubcaption
			\label{fig:350_minority_compared_no_log}
		\end{subfigure}
		\caption{MSD of the center of mass $g_3(t)$ of $N_\text{L} = 350$ chains in bidisperse mixtures as the minority component ($\phi_\text{L} = 0.3$), with $N_\text{S} = 25, 50$, and $100$ chains as the majority component: (a) log-log coordinates; (b) linear coordinates. The MSD of pure $N = 350$ chains is also included for comparison.}	
	\end{figure}

	Effects of varying the short-chain length ($N_\text{S}=25, 50$, and $100$) on the mobility of the longer chain ($N_\text{L} = 350$) are shown in fig.~\ref{fig:350_majority_compared} (long-chain majority with $\phi_\text{L} = 0.7$) and fig.~\ref{fig:350_minority_compared} (long-chain minority with $\phi_\text{L}=0.3$). Panel (b) of both figures replot the data in linear coordinates to highlight
	the changes in diffusion rate.
	At short time, dynamics of the $N=350$ chains stays close to its pure melt limit, but after $t\sim O(10^3)$ ($\sim\tau_\text{e}$), it becomes clear that mixing with shorter chains increases the mobility of the longer chains and the effect is stronger as $N_\text{S}$ decreases. Around $t\sim O(10^4)$ -- i.e., the original $t^{1/2}$ scaling regime in pure entangled melts -- the slope is again raised by the addition of the short chains, which is more clearly observed when the long chains become the minority component (fig. \ref{fig:350_minority_compared}). The onset of this intermediate regime of slower dynamics, which marks $\tau_\text{e}$, is reduced. Overall, the introduction of short chains speeds up the dynamics of the entangled chain species and lessens the extent of entanglement.\par

	\RevisedText{Increasing slope in the $\tau_\text{e}<t<\tau_\text{R}$ segment, with decreasing matrix chain length $N_\text{S}$, was also previously reported by \citet{wang2008constraint} in both the $g_1(t)$ and $g_3(t)$ profiles of their long probe chain.
	For example, in their binary mixtures of $N_\text{L}=350$ and $N_\text{S}=25$ (both semiflexible) chains with $\phi_\text{L}=0.15$, the slope of $g_3(t)$ reached $0.84$, whereas their $N_\text{S}=160$ (and $N_\text{L}=350$) case had a slope of $0.58$.
	The latter is close to our monodisperse $N=350$ case because $N_\text{S}=160$ is already substantially entangled (especially considering their higher chain rigidity which gives lower $N_\text{e}$ than our flexible chains).
	With more short chains in the surrounding (higher $\phi_\text{S}$) or faster relaxation of those chains (lower $N_\text{S}$), CR is stronger, which leads to faster tube relaxation and, eventually, the higher mobility in the longer entangled chains.}

\begin{figure}
		\centering
		\includegraphics[width=3.5in, trim=0 0 0 0, clip]{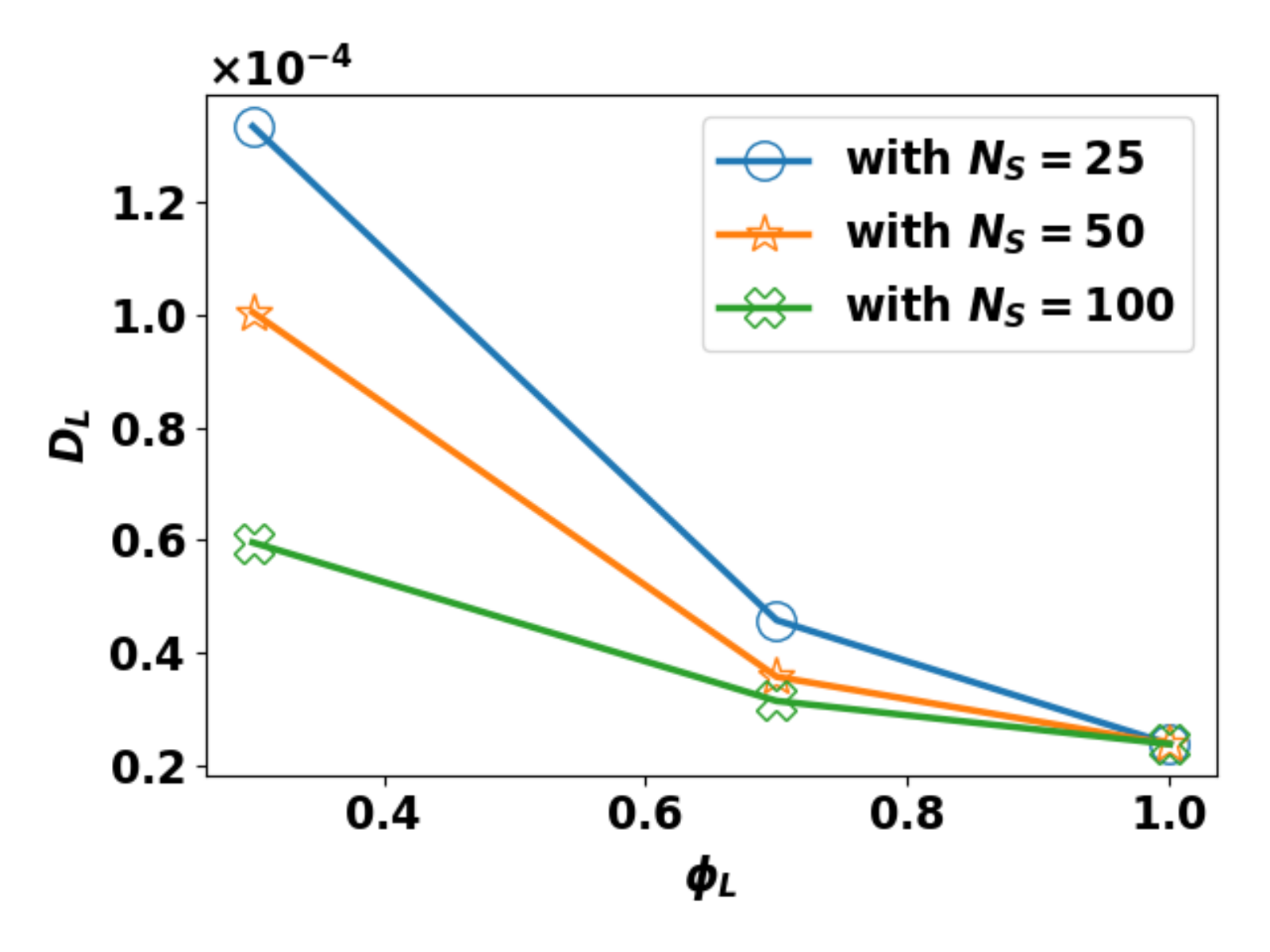}
		\caption{Diffusion coefficient of the long chain ($N_\text{L} = 350$) in bidisperse mixtures with shorter chains of different lengths $N_\text{S}$.}
		\label{fig:diff_long_plot}
	\end{figure}

	The diffusion coefficient of the $N_\text{L}=350$ chains $D_\text{L}$, is again calculated from the MSD data at the long-time limit. The result is plotted in fig. \ref{fig:diff_long_plot}. $D_\text{L}$ increases with the short-chain fraction $\phi_\text{S}=1-\phi_\text{L}$ and the effect is stronger for as $N_\text{S}$ decreases. With $30\%$ short chains in the mixture (i.e., $N_\text{L}=350$ remains the majority), the speed-up is $87\%$, $52\%$ and $34\%$ for $N_\text{S} = 25$, $50$, and $100$ respectively, while when short chains reach $70\%$ (i.e., $N_\text{L}=350$ becomes the minority), the speed-up reaches $455\%$, $320\%$, and $151\%$ for the same three $N_\text{S}$ levels.\par
		
	Speed up of long chain dynamics upon introducing the short chain component \RevisedText{again agrees with various} previous simulation studies\RevisedText{~\citep{Barsky2000,Picu2007a,Lin2007,wang2008constraint,Baig2010}}.
	The results above have also been confirmed experimentally. \citet{wang2004diffusion} studied the effects of short chains on the dynamics of long chains and vice versa using a binary mixture of 1,4-polybutadiene at different weight concentrations of the long chains. The diffusion coefficients of the different chains were measured using pulsed-gradient NMR spin echo measurements. Their results showed that the dynamics of the longer chains were sped up by the shorter chains
	and the effects were stronger
	with decreasing weight fraction of the longer chains. \par

\subsection{Rouse Mode Analysis (RMA)}\label{subsec:RMA}
	MSD analysis above suggests that: (1) dynamics of the short-chain species is impeded by the long chains but still follows similar patterns as the corresponding pure short-chain melt and (2) dynamics of the long-chain species is accelerated by the short chains and features of entanglement become weakened. Here, we apply Rouse mode analysis to directly examine the extent of entanglement effects in different species. We start with a quick review of the Rouse model. It treats the probe chain as a Gaussian chain and considers all its surrounding chains to form a continuous viscous medium -- i.e., a mean-field approach. Dynamics of each bead on the probe chain is described by the inertia-less Langevin equation\cite{doi1987dynamics}. For example, the equation of motion for the $i$-th bead is written as: 
	\begin{equation}\label{eq:eqn_of_motion}
		\zeta\frac{d\vec{r}_i}{dt}=H_\text{s}\left[\left(\vec{r}_{i+1}-\vec{r}_{i}\right)-\left(\vec{r}_{i}-\vec{r}_{i-1}\right)\right]+\vec{f}^\text{r}_{i}
	\end{equation}
	where $\zeta$ is the monomeric friction coefficient, $H_\text{s}$ is the spring constant, $\vec{r}_i$ is the position of the $i$-th bead, and $\vec{f}^\text{r}_{i}$ is the random force exerted on the $i$-th bead
	satisfying
	\begin{equation}
		\left\langle\vec{f}^\text{r}_{i}(t)\vec{f}^\text{r}_{j}(t')\right\rangle = 2\zeta k_\text{B} T\delta_{ij}\delta(t-t')\vec{\vec\delta}
	\end{equation}
	where $\delta(t)$ is the Dirac delta function, $\delta_{ij}$ is the Kronecker delta, and $\vec{\vec\delta}$ is the identity tensor. \Cref{eq:eqn_of_motion} shows that position coordinates of neighboring beads are coupled in their dynamics through spring forces. The RMA projects the original bead coordinates $\vec{r}_{i}$ to a set of mutually orthogonal coordinates known as Rouse modes or normal coordinates $\vec{X}_p$ ($p = 0,1,...,N-1$). We adopt the original form of projection by Rouse\cite{rouse1953theory}
	\begin{equation}\label{eq:rouse_mode}
	\begin{split}
	\vec{X}_p\equiv\sqrt{\frac{2}{N}}\sum_{n=1}^N \vec{r}_i(t)\cos\left(\frac{(i-1/2)p\pi}{N}\right)
	\\(p = 0,1,2,...N-1)
	\end{split}
	\end{equation}
	which is widely used in the literature\cite{verdier1966monte, shaffer1995effects, Kopf1997, vladkov2006linear, kalathi2015rouse}. The $p = 0$ mode describes the motion of the center of mass of the chain and the other modes ($1 \leq p \leq N -1$) describe the internal relaxations of sub-chains, or "blobs", of the size of $N/p$ beads. Each of the transformed coordinate or Rouse mode $\vec{X}_p$ follows Langevin dynamics with its own friction coefficient and random force. Importantly, relaxation of different modes is mutually independent. The autocorrelation function (ACF) of each $p > 1$ mode decays exponentially 
	\begin{equation}
	\left\langle\vec{X}_p(t)\vec{X}_p(0)\right\rangle =
		\left\langle\vec{X}_p^2\right\rangle
		\exp\left(-\frac{t}{\tau_p}\right)
	\label{eq:simple_exponential}
	\end{equation}
	with its own relaxation time $\tau_p$ given by 
	\begin{equation}\label{eq:relax_time}
		\tau_p^{-1} = \frac{12k_\text{B}T}{\zeta b^2}\sin^2\left(\frac{p\pi}{2N}\right)
	\end{equation}
	where $b^2$ is the mean-square bond (spring) length. For leading modes with $p\lesssim N/5$, which describes the motions of larger segments with $N/p \gtrsim 5$ beads, \cref{eq:relax_time} can be approximated by \cite{rouse1953theory}
	\begin{gather}\label{eq:approx_relax_time}
		\tau_p=\frac{\zeta b^2}{3\pi^2 k_\text{B}T}\left(\frac{N}{p}\right)^2.
	\end{gather} 
	
	Rouse model is commonly used to describe the dynamics of unentangled polymer melts. With increasing chain length, topological constraints set in and relaxation dynamics changes.
	For entangled chains, we may still project the coordinates to $\vec X_p$ using \cref{eq:rouse_mode} but the ACFs no longer follow simple exponential decay. A stretched exponential is often used instead \cite{Kopf1997, shaffer1995effects, padding2001uncrossability,li2012nanoparticle, Kalathi2014} 
	\begin{equation}
		\left\langle\vec{X}_p(t)\vec{X}_p(0)\right\rangle=
			\left\langle\vec{X}_p^2\right\rangle
			\exp\left[-\left(\frac{t}{\tau_p^*}\right)^{\beta_p}\right]
		\label{eq:stretched_exponential}
	\end{equation}
	where $\tau_p^*$ and $\beta_p$ are the time scale and exponent (stretching parameter) for the $p$-th mode. The relaxation time of a stretched exponential can be defined as
	\begin{equation} \label{eq:stretched_relaxation_time}
		\tau_p\equiv\int_0^\infty
		\exp\left[-\left(\frac{t}{\tau_p^*}\right)^{\beta_p}\right]dt =
		\left(\frac{\tau_p^*}{\beta_p}\right)
		\Gamma\left(\frac{1}{\beta_p}\right)	
	\end{equation}
	where $\Gamma(x)$ is the gamma function. Note that at the simple exponential limit, i.e., $\beta_p\to1$, the two time scales are the same $\tau_p=\tau_p^*$.

\begin{figure}
		\centering
		\includegraphics[width=3.5in, trim=0 0 0 0, clip]{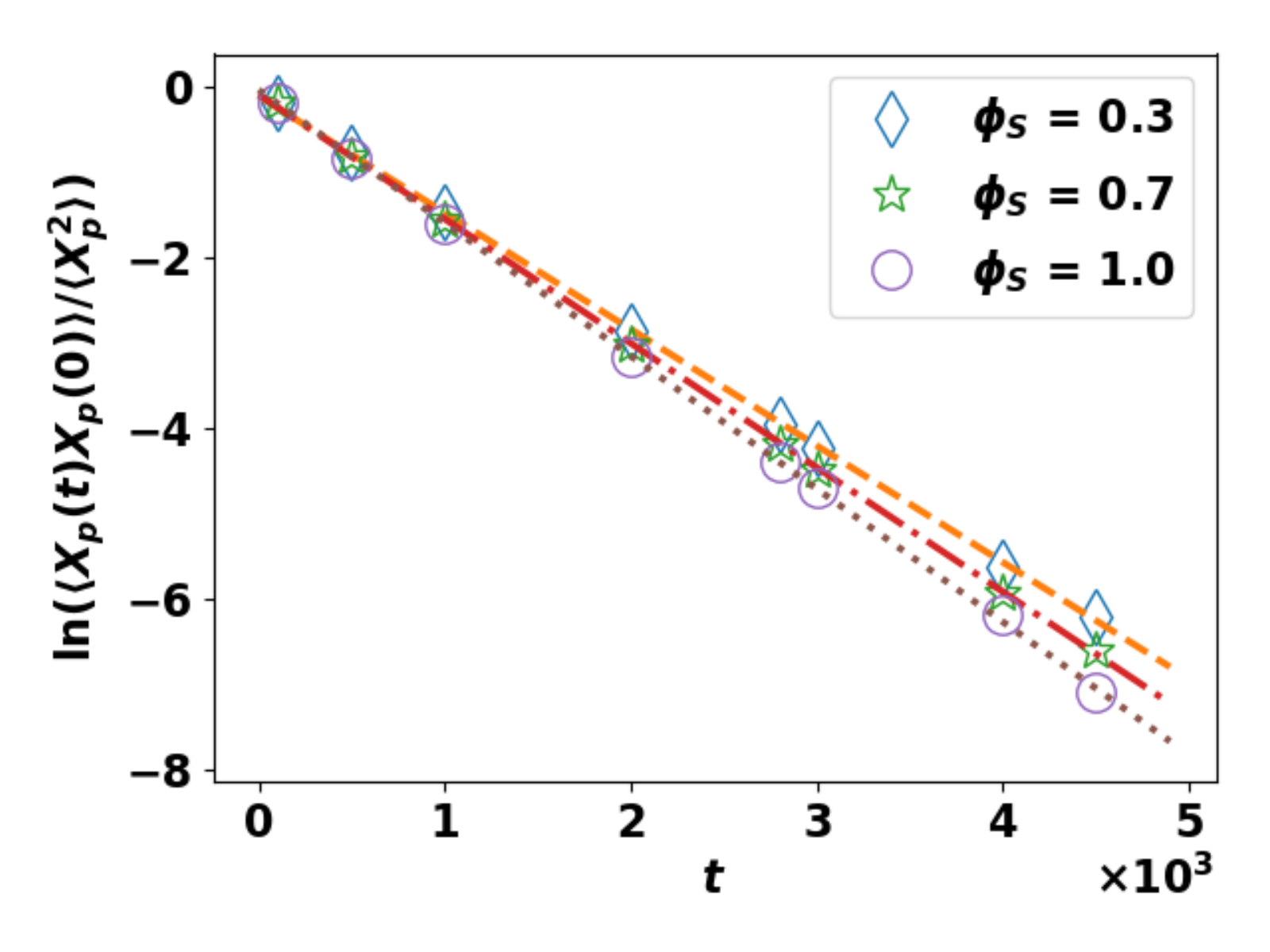}
		\caption{Relaxation of the $p=1$ Rouse mode of $N_\text{S}=25$ chains in its pure melt and in bidisperse mixtures with $N_\text{L}=350$ chains. Lines represent linear regression corresponding to the simple exponential relaxation of \cref{eq:simple_exponential}.}
		\label{fig:25_RouseMode_pure}
	\end{figure}

We start with the relaxation of $N_\text{S} = 25$ chains in its pure melt ($\phi_\text{S}=1.0$)  and in mixtures with $N_\text{L}=350$ chains. Figure \ref{fig:25_RouseMode_pure} shows the time ACF of the $p = 1$ mode at different mass fractions. In all cases, the logarithm of the ACF follows a straight line for nearly the whole time range -- i.e., the simple exponential decay as given in \cref{eq:simple_exponential} accurately describes the relaxation dynamics of short chains. Relaxation dynamics of unentangled monodisperse melts of KG chains is known to be well approximated by the Rouse model despite its many simplifications \cite{Kremer1990}. However, we find that even in mixtures with a long-chain species well beyond the entanglement threshold, the nature of the short-chain dynamics is not changed at least for $\phi_\text{S}$ down to $30\%$.
(At very low $\phi_\text{S}$, we do expect the dynamic pattern to differ -- indeed, diffusion of small molecules, at extremely low concentration, in a matrix of long-chain polymers is known to display jerky ``hop''-like movements~\citep{Xi_Trout_JPCB2013}.)
The presence of longer chains seemingly do not have any effect other than to increase the relaxation time of the shorter chains -- note in fig. \ref{fig:25_RouseMode_pure} that the pure $N_\text{S}=25$ melt has the steepest slope and the relaxation slows down with increasing long chain fraction ($\phi_\text{S}$ decreases). This is consistent with the earlier discussion that having longer chains in the background medium increases the effective friction coefficient of the short chains. The results for $N_\text{S}$ = 50 are similar and thus not shown here.\par
	
	\begin{figure}
		\centering
		\includegraphics[width=3.5in, trim=0 0 0 0, clip]{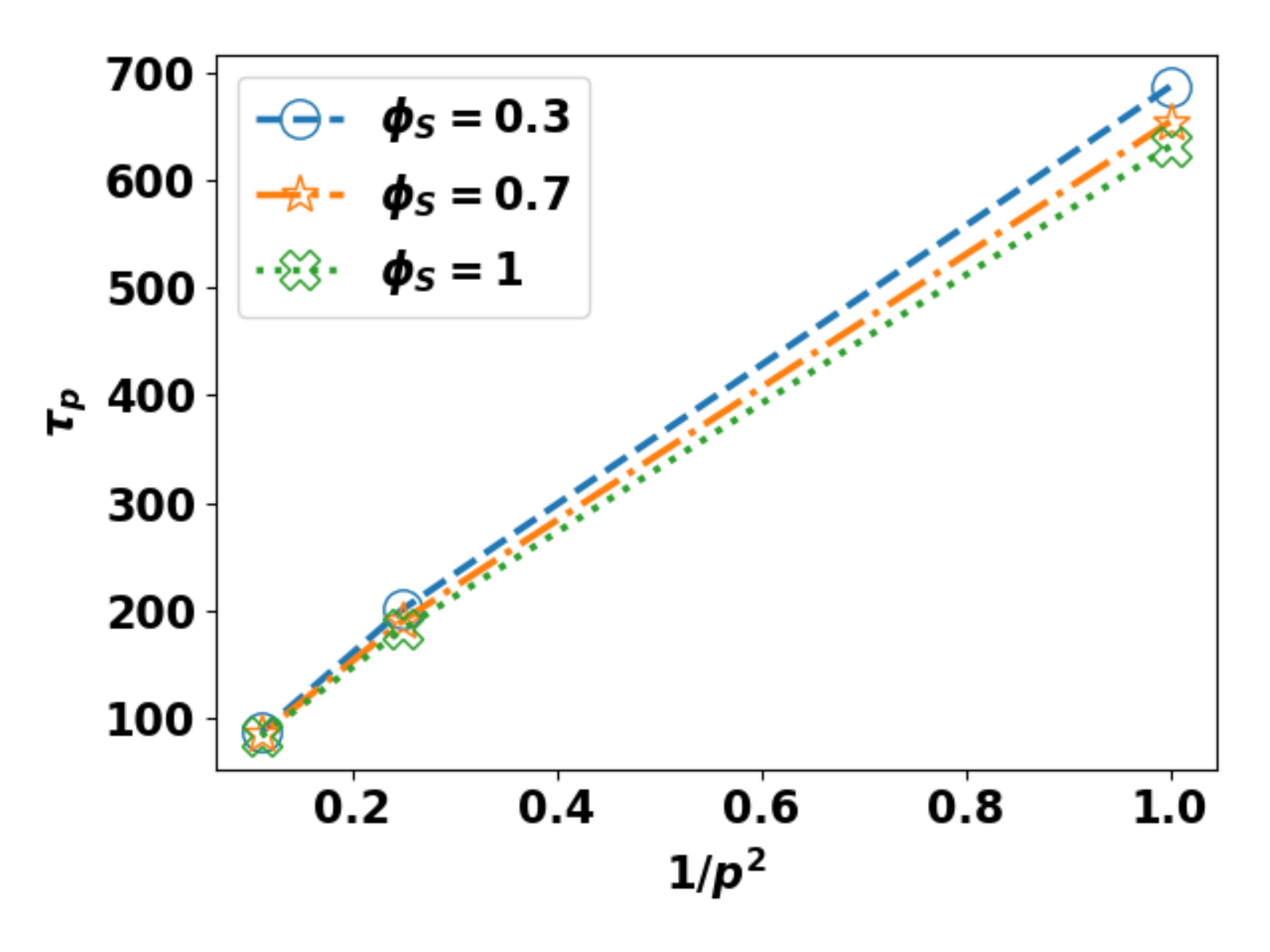}
		\caption{Single-exponential relaxation time $\tau_p$ of $N_\text{S}=25$ chains in its pure melt ($\phi_\text{S}=1.0$) and in bidisperse mixtures with $N_\text{L}=350$ chains ($p=1$ to $p=3$).}
		\label{fig:25_relax_time}
	\end{figure}		

	\begin{figure}
		\centering
		\includegraphics[width=3.5in, trim=0 0 0 0, clip]{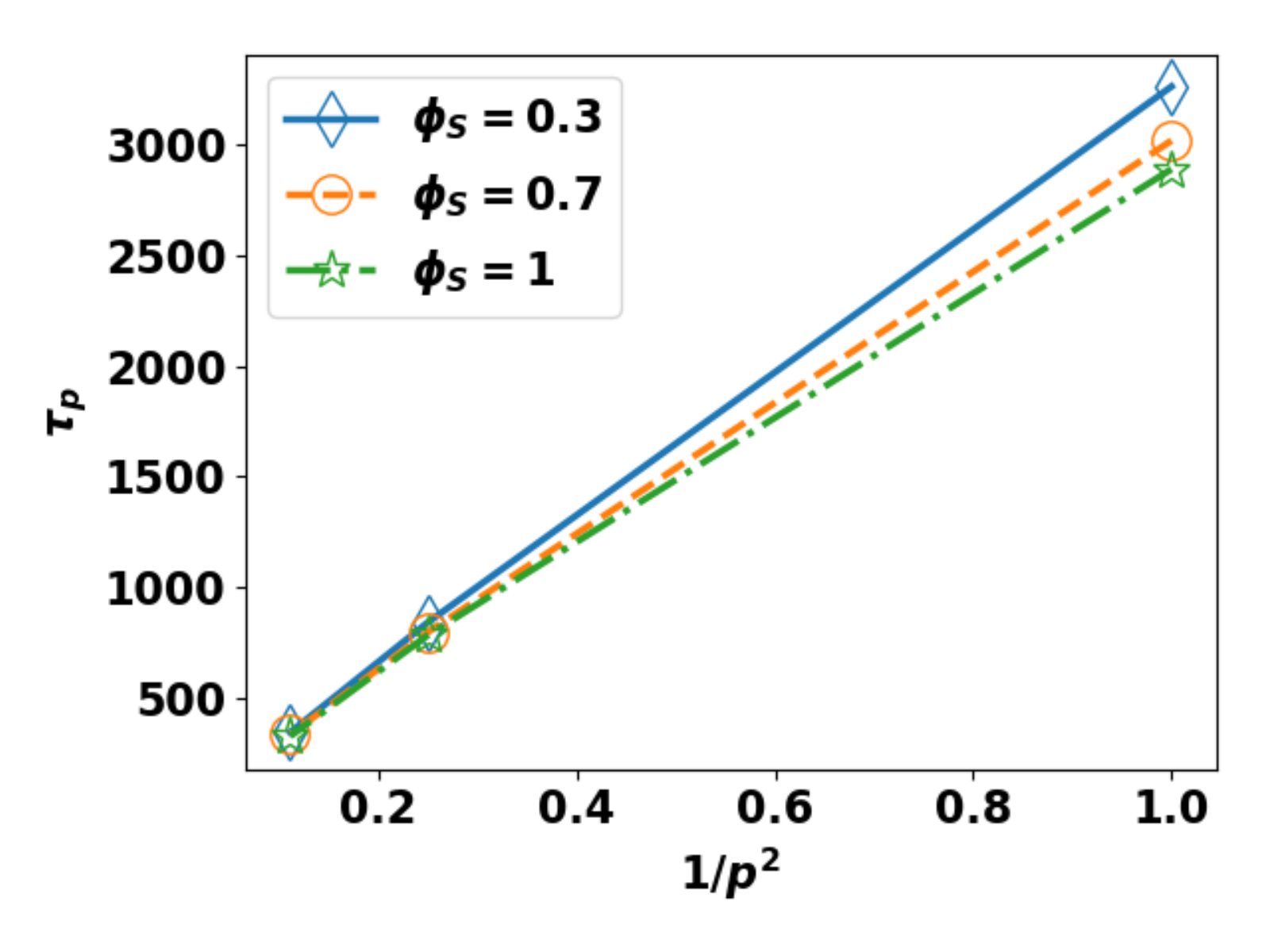}
		\caption{Single-exponential relaxation time $\tau_p$ of $N_\text{S}=50$ chains in its pure melt ($\phi_\text{S}=1.0$) and in bidisperse mixtures with $N_\text{L}=350$ chains ($p=1$ to $p=3$).}
		\label{fig:50_relax_time}
	\end{figure}		

	Fitting the ACF to the single exponential of \cref{eq:simple_exponential} provides the relaxation time, which is plotted against $1/p^2$ in Figures \ref{fig:25_relax_time} and \ref{fig:50_relax_time} for $N_\text{S} = 25$ and $50$ chains respectively. From \cref{eq:approx_relax_time}, for the Rouse model, $\tau_p$ versus $1/p^2$ should give a straight line, at least for small $p$, whose slope equals the longest relaxation time $\tau_1$. For $N_\text{S}=25$ (\cref{fig:25_relax_time}), this Rouse behavior is clearly demonstrated, whereas some small deviations are observed in $N_\text{S} = 50$ case (\cref{fig:50_relax_time}). Introducing a longer $N_\text{L}=350$ species does not change the nature of the dynamics, but the relaxation time of the shorter chains increases with increasing long-chain content. In both figures, the highest $p$ mode available is limited by the sampling frequency -- the frequency at which coordinates were stored -- used in our simulation.\par

\RevisedText{Slowdown of short-chain dynamics in the presence of longer chains in its surroundings is intuitively predictable and well established in the literature~\citep{Barsky2000,Picu2007a,wang2008constraint,Baig2010}.
Nevertheless, our finding, from results here as well as back in \cref{fig:diff_short_N}, that the reduced mobility can be fully described by an increased friction factor, which is only a function of composition $\phi_\text{L}$ and does not vary with $N_\text{S}$, has never been reported before to our best knowledge.
\citet{Kopf1997} found that when two unentangled isotope chain species with identical chain length but different monomeric mass are mixed, they retain the same Rouse dynamics as their pure melts but with different effective monomeric friction factors -- the light and heavy components see their friction factor to increase and decrease, respectively.
The effective friction factor depends on both the volume fraction of heavy chains and the mass ratio between the components. However, at least for $10\leq N\leq 30$ investigated in that study, it does not depend on the chain length $N$.
What we find here is that, for a short unentangled chain species, mixing with a longer species, which is well within the entangled regime, does not change its own dynamical patterns, even when, e.g., in the case of $N_\text{S}=50$, its own length already exceeds $N_\text{e}$.}

\begin{figure}
		\centering
		\includegraphics[width=3.5in, trim=0 0 0 0, clip]{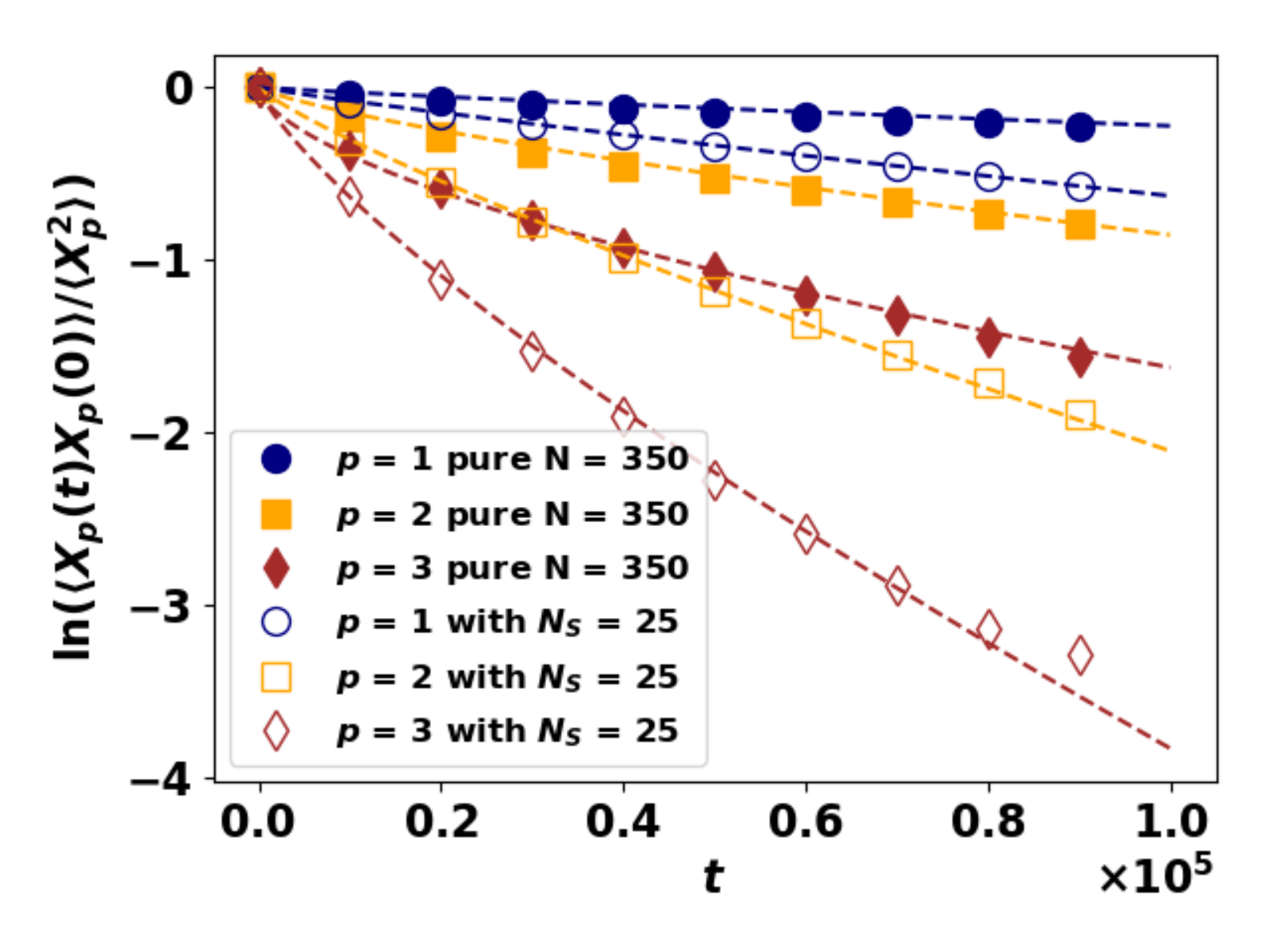}
		\caption{Relaxation of the leading Rouse modes of $N=350$ chains in its pure melt (filled symbols) and as the minority component in a bidisperse mixture with $N_\text{S}=25$ (empty symbols; $\phi_\text{L}=0.3$). Lines represent stretched exponential fits (\cref{eq:stretched_exponential}).}
		\label{fig:log_auto_t}
	\end{figure}

	Departure from Rouse dynamics is observed as chain length gets longer and entanglement effects set in. Rouse mode  projection (\cref{eq:rouse_mode}) may still be applied to longer chains, but relaxation of each mode is no longer independent. As a result, its ACF does not follow the simple exponential decay as given in \cref{eq:simple_exponential}. \Cref{fig:log_auto_t} shows stretched exponential (\cref{eq:stretched_exponential}) fits to the ACFs of leading Rouse modes of the monodisperse $N=350$ system as well as its mixture with short $N_\text{S} = 25$ chains as the diluent. In both cases, it appears that the $p=1$ mode may still be reasonably approximated by a simple exponential (i.e., close to straight lines in the figure). However, at $p=3$, curvature in the profile is too strong to be ignored and a stretched exponential is required. The same behavior is observed in higher modes such as $p=5$ and $7$ (not shown here). Coupling between different modes is attributed to the topological constraints imposed by entanglements. Indeed, using a simple lattice model, Shaffer \cite{shaffer1995effects} showed that allowing chains to cross one another would recover the Rouse dynamics in otherwise entangled chains. For bidisperse mixtures, we observe that introducing short $N_\text{S}=25$ chains as a diluent, at least for $\phi_{S}$ up to $70\%$ shown in \cref{fig:log_auto_t}, does not eliminate this non-Rouse behavior, even though the relaxation dynamics of the $N_\text{L}=350$ chains is significantly accelerated (compared with its pure melt). \par

\begin{figure}
		\centering
		\begin{subfigure}[b]{0.49\textwidth}
			\centering
			\includegraphics[width=\textwidth]{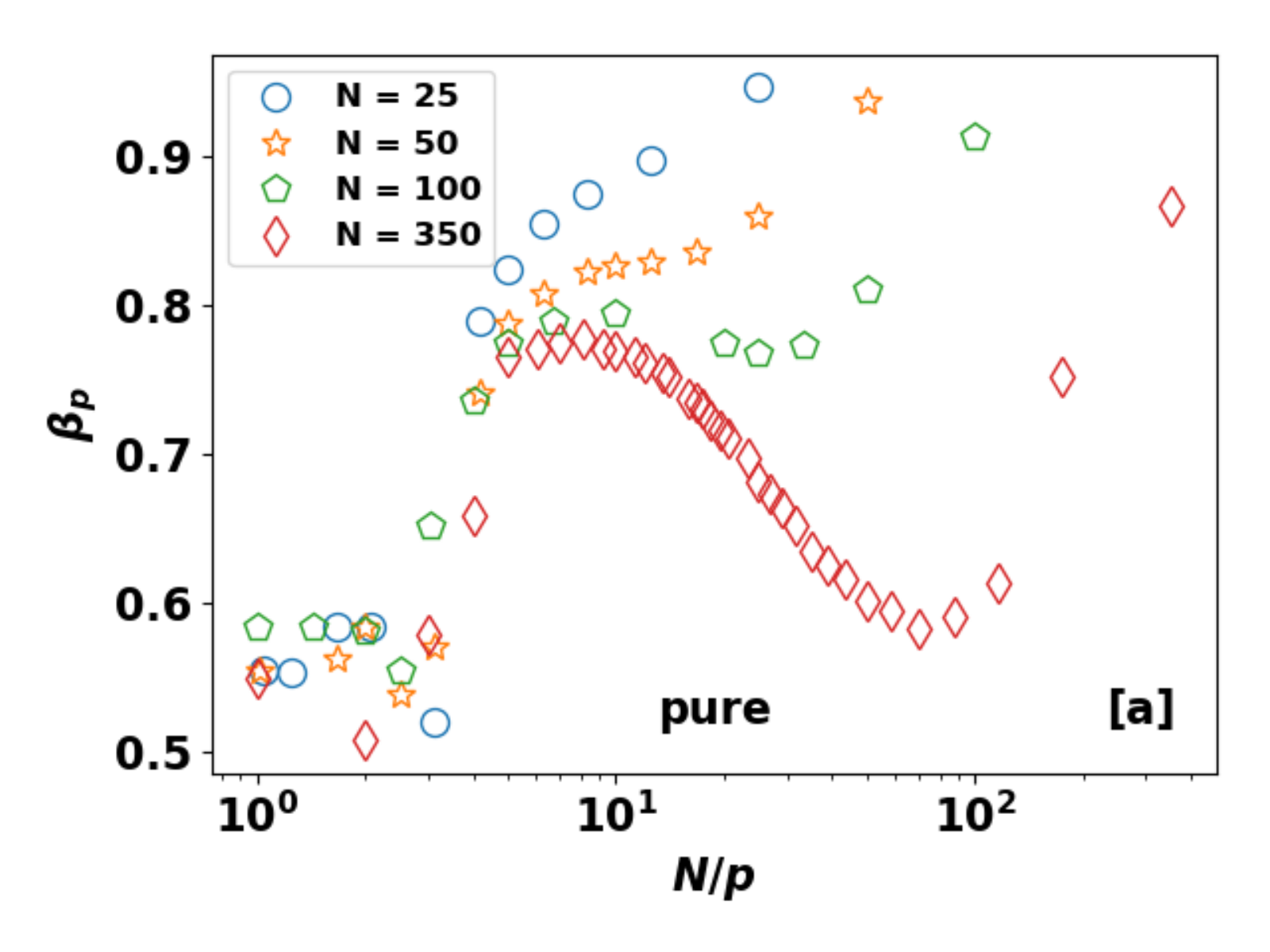}
			\phantomsubcaption
			\label{fig:monodisperse_beta}
		\end{subfigure}
		\begin{subfigure}[b]{0.49\textwidth}
			\centering
			\includegraphics[width=\textwidth]{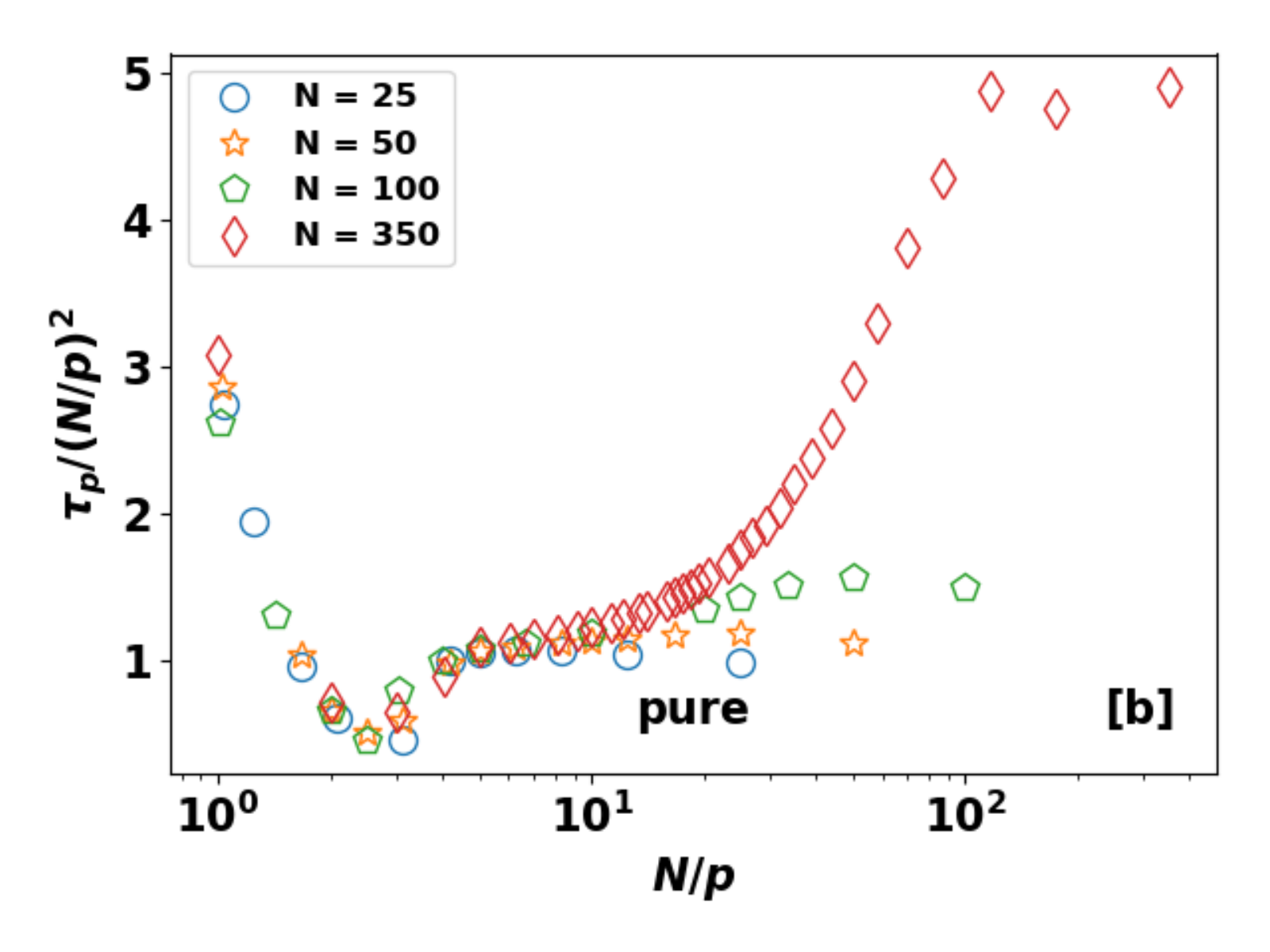}
			\phantomsubcaption
			\label{fig:monodisperse_relaxation_time}
		\end{subfigure}
		\begin{subfigure}[b]{0.49\textwidth}
			\centering
			\includegraphics[width=\textwidth]{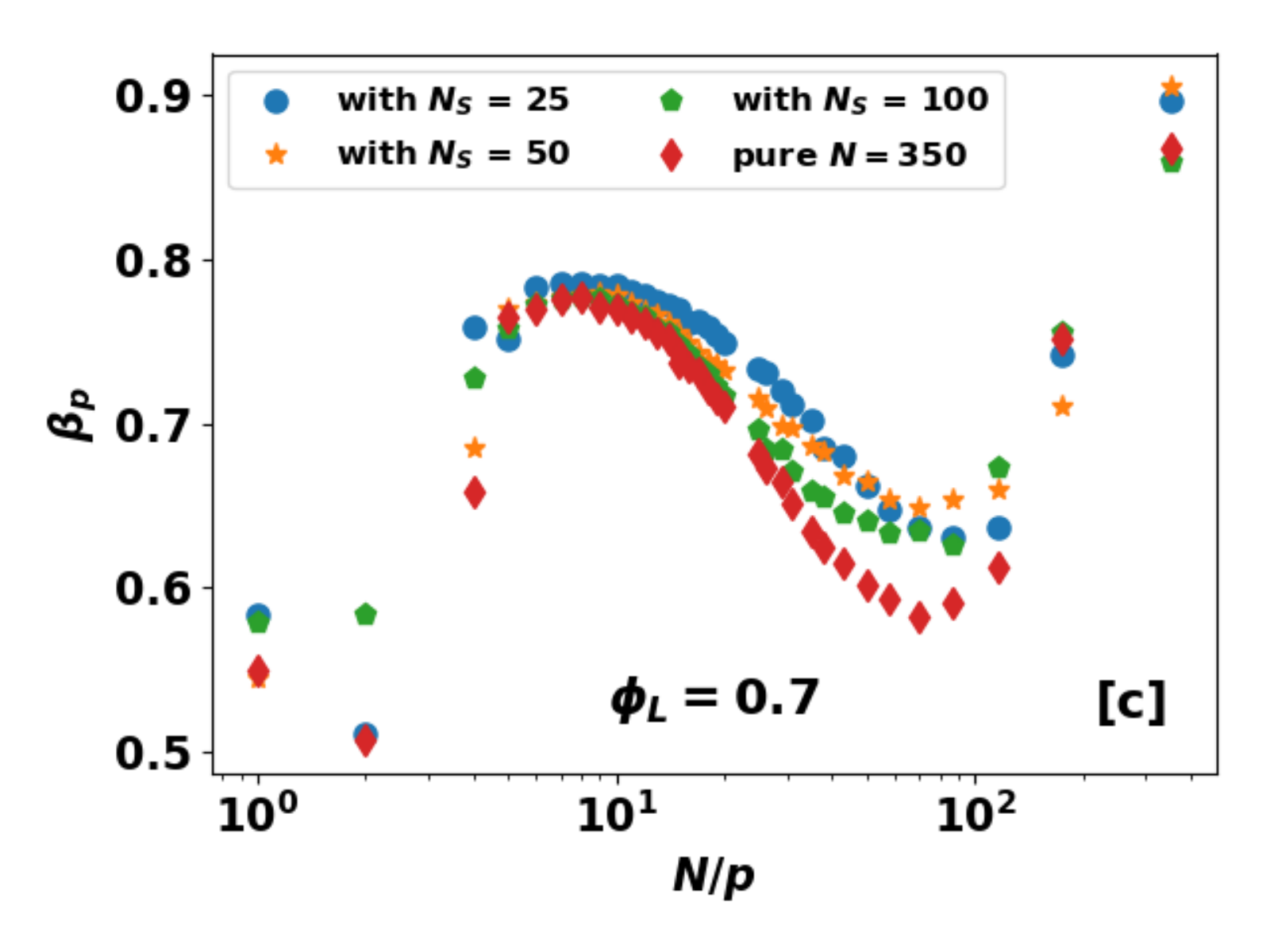}
			\phantomsubcaption
			\label{fig:350_majority_beta}
		\end{subfigure}
		\begin{subfigure}[b]{0.49\textwidth}
			\centering
			\includegraphics[width=\textwidth]{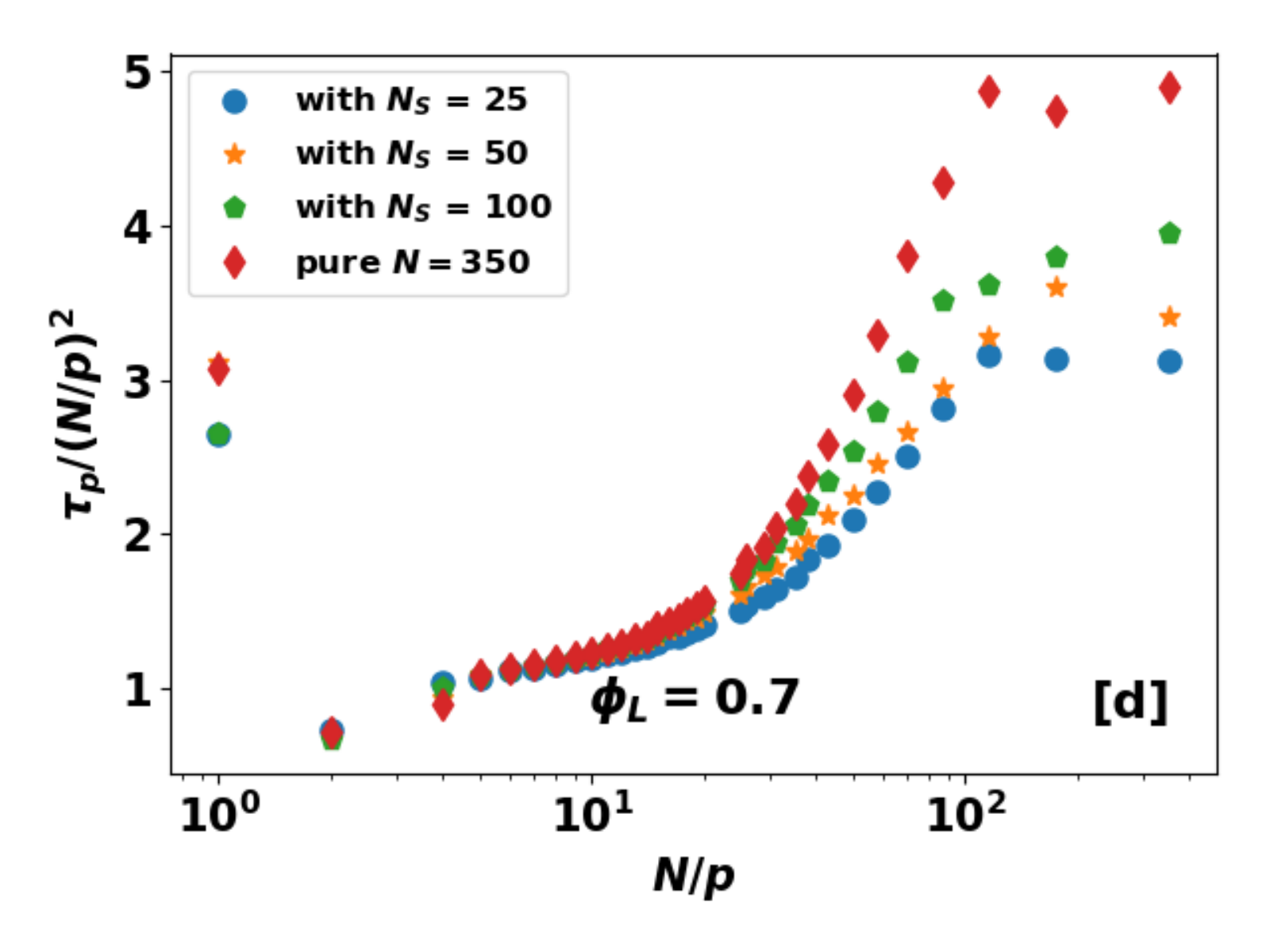}
			\phantomsubcaption 
			\label{fig:350_rousemajority_time}
		\end{subfigure}
		\begin{subfigure}[b]{0.49\textwidth}
			\centering
			\includegraphics[width=\textwidth]{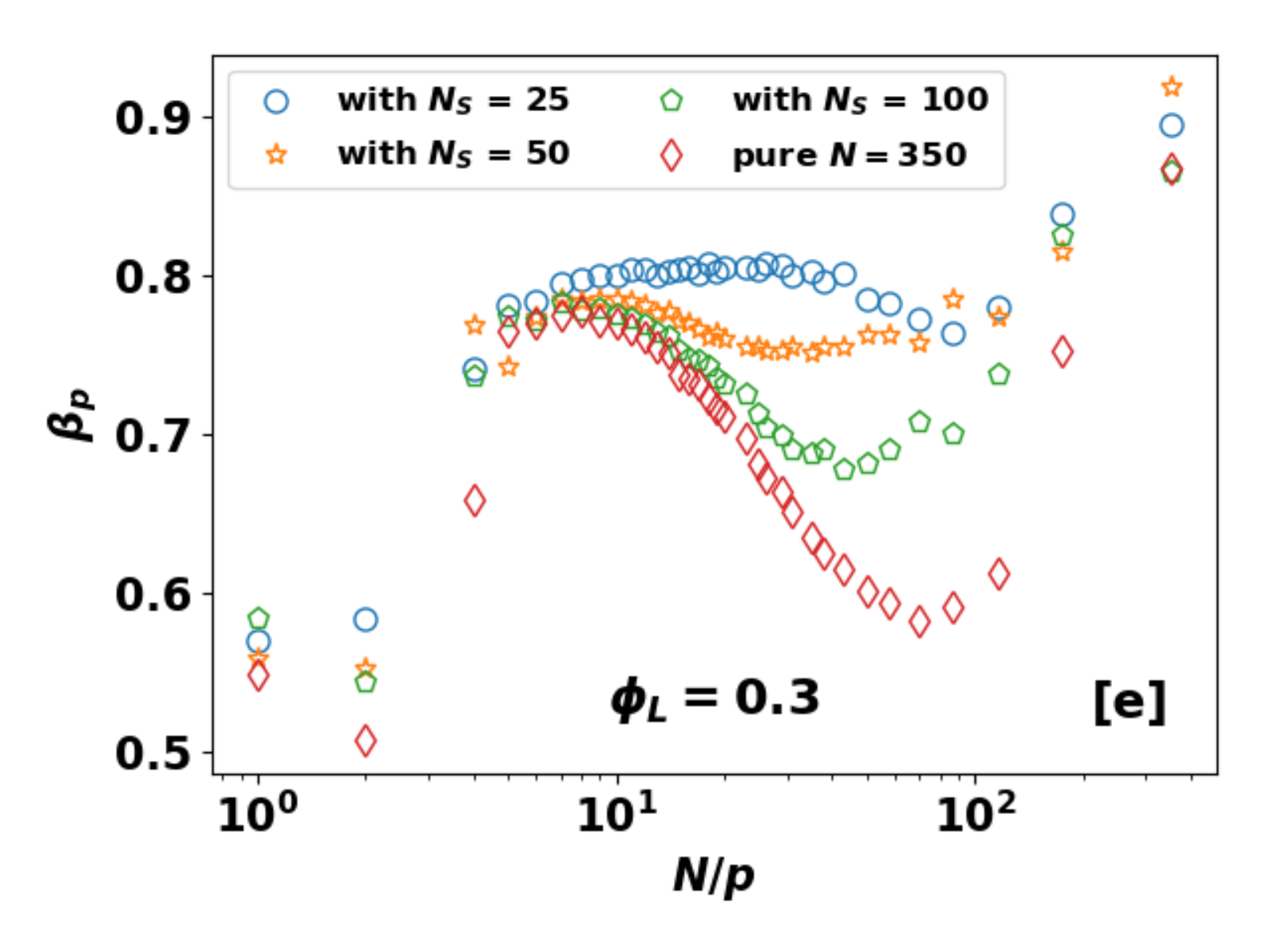}
			\phantomsubcaption
			\label{fig:350_minority_beta}
		\end{subfigure}
		\begin{subfigure}[b]{0.49\textwidth}
			\centering
			\includegraphics[width=\textwidth]{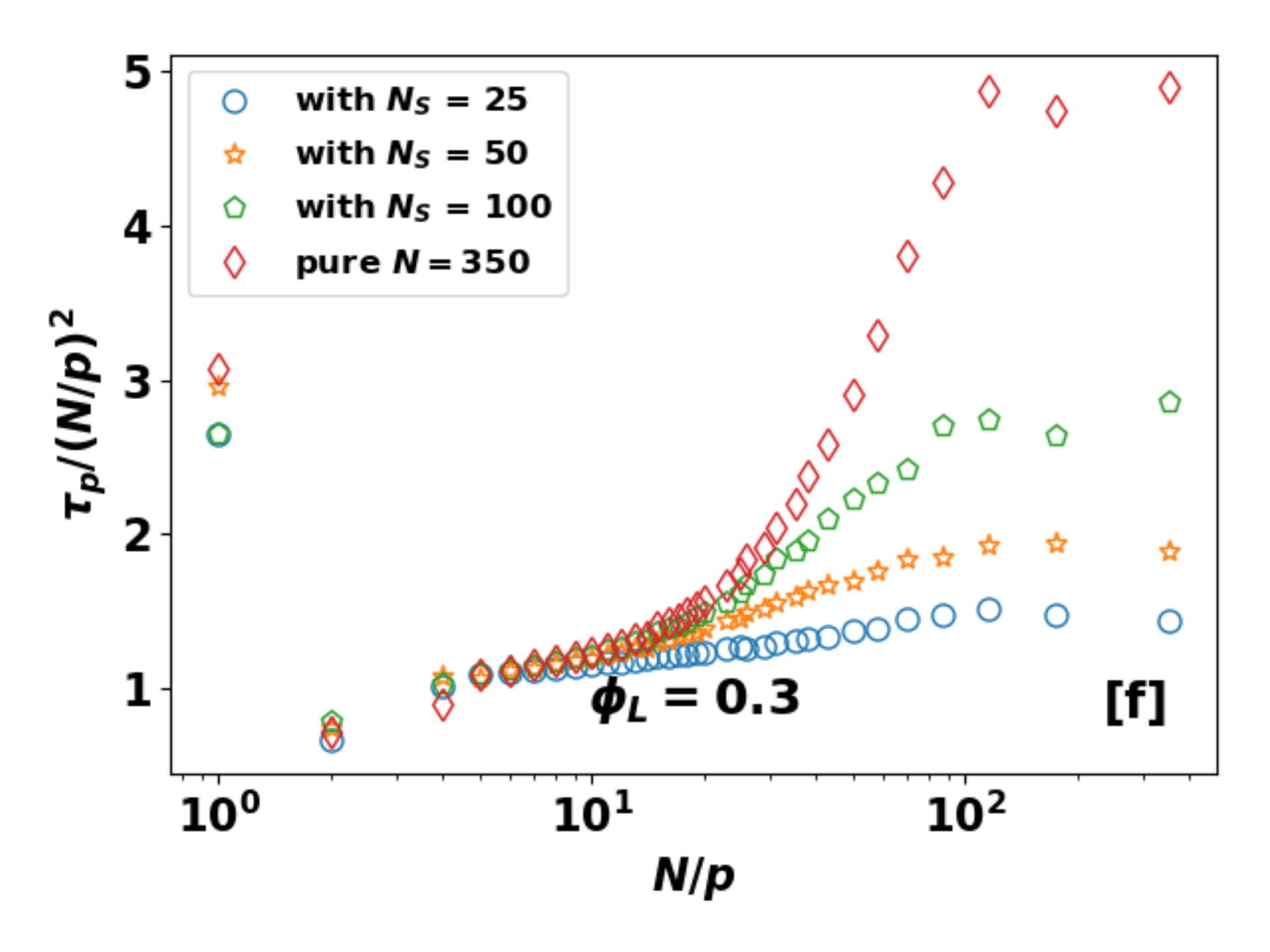}
			\phantomsubcaption 
			\label{fig:350_rouseminority_time}
		\end{subfigure}		
		\caption{Rouse mode analysis with stretched exponential fitting -- stretching parameter $\beta_p$ ((a),(c), and (e)) and relaxation time $\tau_p$ (scaled by $(N/p)^2$; (b),(d), and (f)) -- of monodisperse melts of varying chain length ((a) and (b)) and
		\RevisedText{the $N_\text{L}=350$ component in bidisperse mixtures with the longer chains}
		as the majority ($\phi_\text{L}=0.7$; (c) and (d)) and minority ($\phi_\text{L}=0.3$; (e) and (f)) component.}
	\end{figure}

	Curvature in the $\ln(\langle\vec{X}_p(t)\vec{X}_p(0)\rangle/\langle\vec X_p^2\rangle)$ vs $t$ plot is measured by the stretching parameter $\beta_p$, which can thus be viewed as an indicator of non-Rouse behaviors ($\beta_p =1$ in the purely Rouse limit). Figure \ref{fig:monodisperse_beta} shows $\beta_p$ as a function of $N/p$ (which measures the number of beads in each segment or sub-chain described by the $p$-th mode) for monodisperse melts of different chain lengths. At the high-$p$ (small $N/p$) end, all curves approximately overlap regardless of the chain length (entangled or not), indicating that relaxation of small segments is independent of the overall chain length. This terminal $\beta_p$ value of $0.5\sim 0.6$ is significantly lower than 1 -- departure of small-scale segmental relaxation from the Rouse model is obvious even for the shortest chains. This is likely due to the differences between the KG chain used in our simulation and the Gaussian chain in the Rouse model. In particular, the latter does not consider the excluded-volume effect between beads, which is more important in dynamics at small scales.
\RevisedText{Indeed,
	strong departure of $\beta_p$ from $1$ at small scales was noted in several earlier studies~\citep{padding2002time, Lin2007,Kalathi2014}.}
	At $N/p\approx3$, $\beta_p$ starts to rise steeply, reaching nearly $0.8$ at $N/p \approx 5$ (segment size of 5 beads), after which curves of different $N$ separate. For $N=25$ (where $N/p=5$ corresponds to the $p=5$ mode), $\beta_p$ continues to increase with lowering $p$, ending well above $0.9$ for the $p=1$ mode. For $N=50$, $\beta_p$ plateaus around $0.8$ until $p=3$ mode after which it again quickly rises above 0.9.\par

	Entanglement effects are most clearly seen in the $N=350$ case where, after the plateau at $5\lesssim N/p\lesssim 10$, $\beta_p$ quickly declines and reaches its minimum at $N/p\sim 70$. This minimum $\beta_p$ is around $0.6$. Strong departure from the Rouse model at this length scale is attributed to the topological constraints between entangled chains: each chain is now constrained to its surrounding tube (the chain follows reptation motion) and can no longer meander freely in the three-dimensional space. The size of segments $N/p$ strongly affected by this effect also coincides with the entanglement strand size $N_\text{e}$ of the KG model ($30\sim 80$).
	Interestingly, leading modes ($p=4$ to $1$) again show a rapid surge of $\beta_p$ -- relaxation of largest segments returns to the Rouse-like single exponential behavior, which is consistent with the observation in fig. \ref{fig:log_auto_t}
	that a simple exponential adequately captures the relaxation behavior of the $p=1$ mode. One can rationalize this considering that at length scales $\gg N_\text{e}$, \RevisedText{the conformation of} the constraining tube itself
	\RevisedText{(or, more accurately, the primitive path)}
	undergoes multiple turns. In addition, over the time scale of $\tau_\text{d}$ (the longest relaxation time of entangled chains),
	\RevisedText{surrounding chains all have sufficient time to relax, which causes significant CR, and the tube (primitive path) conformation changes substantially through its own Rouse motion~\citep{viovy1991constraint}.}
	Chain motion at the largest scales is thus again described by a three-dimensional random walk.
	The $N=100$ case is similar to $N=50$ but a small dip in $\beta_p$ is found near $N/p \approx 30$ as a sign of weak entanglement.

	Dependence of relaxation time $\tau_p$, as calculated from \cref{eq:stretched_relaxation_time}, on the segment size $N/p$ is shown in \cref{fig:monodisperse_relaxation_time} (again for monodisperse melts). Rouse model predicts (\cref{eq:approx_relax_time}) $\tau_p/(N/p)^2$ to be a constant for $N/p\gtrsim5$ which is indeed observed in the $N=25$ and $50$ cases in our simulation. Departure from the plateau is found at smaller scales ($N/p \lesssim$ 5), where the relaxation times of all chain lengths again overlap. Entanglement causes a sudden slowdown in the dynamics, which shows as a surge in $\tau_p/(N/p)^2$ starting at $N/p\sim\mathcal{O}(10)$. At the long-segment (small-$p$) limit, another, much higher, plateau is reached. This conforms with the tube model prediction \cite{doi1988theory} of 
	\begin{gather}
		\tau_p\sim\left(\frac{N}{N_\text{e}}\right)\left(\frac{N}{p}\right)^2.
	\end{gather}
	A raised right-end plateau is also discernible in the $N=100$ case, although the level of elevation (from the Rouse plateau) is small, indicating that $N=100$ is close to the onset of entanglement.\par

	\RevisedText{For monodisperse melts, stretched exponential analysis of Rouse modes has been performed in a number of previous studies using various models.
	\citet{shaffer1995effects}'s BFM, despite its many differences in model construction with the KG model used in our study, showed strikingly similar results.
	Their $\beta_p$ profile started with a plateau value between $0.7$ and $0.8$, which agrees with our plateau at $5\lesssim N/p\lesssim 10$.
	They did not report a lower $\beta_p$ level for smallest $N/p$ but such discrepancy at smallest scales is expected given the differences between the models.
	For entangled chains, they also showed a deep dip at higher $N/p$, occurring also at $N/p\approx 60\sim70$ for their longest chains ($N=300$ and $500$). 
	For reference, their model reported $N_\text{e}\approx 32$ based on the self-diffusion coefficient of chains -- i.e., their dip also occurred at $N/p\approx 2N_\text{e}$, which was quantitatively consistent with our observation.
	Their $N=160$ chains, similar to our $N=100$ case, showed a shallower dip occurring at smaller $N/p$, which is a sign of weak entanglement expected in the transitional regime.
	Their $\tau_p/(N/p)^2$ profiles also showed a raised plateau at the high-$N/p$ end for entangled chains.
	Interestingly, both hallmarks of entanglement (dip in $\beta_p$ and raised plateau in $\tau_p/(N/p)^2$) disappeared if chains were allowed to cross.
	More recent studies based on the KG model, for $N=\;$\numrange{500}{2000}, also showed an ``entanglement dip'' at $N/p\approx 70\sim 100$ with $\beta_p$ dropping down to $\sim 0.5$~\citep{Kalathi2014,li2012nanoparticle,Hsu2017}.
	On the other hand, the diamond lattice model by \citet{Lin2007} showed $\beta_p$ to increase monotonically from $\sim 0.4$ to a plateau of $\sim 0.7$ at the high-$N/p$ end without the ``entanglement dip''.
	Similarly, \citet{padding2002time} performed MD using a coarse-grained polyethylene model and also only reported a plateau at the high-$N/p$ end for highly entangled chains.
	Note that unlike the KG model, in which the LJ potential of tightly-bonded beads is sufficient to prevent chain crossing, non-crossability had to be explicitly enforced in the above two models.
	For \citet{padding2002time}, the coarse-grained non-bonded interactions were too soft to prevent chain crossing and an additional bond-crossing potential was imposed  (the so-called TWENTANGLEMENT algorithm~\citep{padding2001uncrossability}).
	Thus, whether or not an ``entanglement dip'' would occur seems to depend on the specific treatment of non-crossability.
	}\par

	In binary mixtures with shorter chains where the $\RevisedText{N_\text{L}}=350$ species remains the majority component (figs. \ref{fig:350_majority_beta} and \ref{fig:350_rousemajority_time}), the same hallmarks of entanglement -- i.e., dip in $\beta_p$ at $\mathcal{O}(10)\lesssim N/p \lesssim \mathcal{O}{(10^2)}$ and raised plateau in $\tau_p/(N/p)^2$ at $N/p\gtrsim\mathcal{O}(10^2)$ -- are preserved. Introduction of the short chain diluent lessens the extent of entanglement, as reflected by the shallowing of the $\beta_p$ dip and reduction in the \RevisedText{raised} $\tau_p/(N/P)^2$ plateau magnitude.
\RevisedText{At its core, this is still a CR effect -- faster relaxation of the shorter chains in its surroundings emancipates the long chain from topological constraints earlier.
This can be described as faster tube Rouse motion~\citep{viovy1991constraint}, but as the diluent chains get shorter, tube dilation~\citep{doi1987dynamics} also plays a role.
Indeed, \citet{Baig2010} showed that tube dilation only occurs when $N_\text{S}<N_\text{e}$, whereas for larger $N_\text{S}$, the diluent accelerates tube relaxation without changing its diameter.}

	The effect strengthens as the diluent chain length decreases. Increasing the short-chain mass fraction to $70\%$ (figs.~\ref{fig:350_minority_beta} and \ref{fig:350_rouseminority_time}) significantly alleviates entanglement and, with $N_\text{S}=25$ as the diluent, dynamics of $\RevisedText{N_\text{L}}=350$ chains is pushed to the marginally entangled limit (similar to the pure $N=100$ case in figs.~\ref{fig:monodisperse_beta} and \ref{fig:monodisperse_relaxation_time}).
	\RevisedText{Increasing CR with increasing $\phi_\text{S}$ and with decreasing $N_\text{S}$ are both expected and well established~\citep{wang2008constraint,Lin2007,Baig2010}.}

\subsection{Stress relaxation}\label{subsec:rheology}
	Previous sections focused on the dynamics of individual chains -- in the case of bidisperse mixtures, discussion of MSD and RMA shows how the relaxation of one chain type is affected by the dynamics of the other. We turn now to the stress relaxation dynamics of the melt in its entirety.
	Stress relaxation modulus $G(t)$ is defined as the ratio of the time-dependent shear stress following a small step strain to the strain magnitude. $G(t)$ contains the full specrum of information about the material linear viscoelasticity and is sensitive to the MWD of the polydisperse mixture. A bidisperse mixture offers a simple model for studying the effects of chain-length disparity on $G(t)$, which will further contribute to the general understanding of MWD effects on polymer rheology.
	
	In equilibrium molecular dynamics, the Green-Kubo relation relates $G(t)$ to the ACF of shear stress fluctuations:
	\begin{equation}
		G(t)= \frac{V}{k_\text{B} T}\left\langle\sigma_{xy}(t_0)\sigma_{xy}(t_0+t) \right\rangle
	\end{equation}
	where $V$ is the volume of the system, $T$ is the temperature, and $\sigma_{xy}$ is the instantaneous shear stress. The stress relaxation modulus is rather difficult to calculate due to intense stress fluctuations that are intrinsic to small-scale systems, which has a particularly strong impact on the stress ACF at the terminal (long-time) regime. In isotropic fluids, stress ACFs in multiple directions can be averaged in an attempt to reduce fluctuations\cite{daivis1994comparison}. We use the particular form of 
	\begin{gather}
	\begin{split}
		G(t) &= \frac{V}{5k_BT}\left[
			\langle\sigma_{xy}(t)\sigma_{xy}(0)\rangle +
			\langle\sigma_{yz}(t)\sigma_{yz}(0)\rangle
		\right.\\&\left.+
			\langle\sigma_{zx}(t)\sigma_{zx}(0)\rangle
		\right]
		+\frac{V}{30k_BT}\left[
			\langle N_{xy}(t)N_{xy}(0)\rangle
		\right.\\&\left.+
			\langle N_{xz}(t)N_{xz}(0)\rangle +
			\langle N_{yz}(t)N_{yz}(0)\rangle
		\right]
	\end{split}
	\end{gather}
	where
	\begin{gather}
		N_{\alpha \beta}=\sigma_{\alpha \alpha}-\sigma_{\beta \beta}.
	\end{gather}
	The same expression was also used in \citet{Ramirez2010}. The number of different components for averaging is nevertheless still too small to significantly reduce the noise in the signal \cite{xi2019molecular}. We further used the multi-tau correlator method developed by \citet{Ramirez2010} which uses a hierarchical multi-level data structure to store and filter time series and calculate correlation functions on the fly. In its data structure, level 0 stores the most recent $p$ data points in the time series. At level 1, the first entry stores the average value of the most recent $m$ ($m<p$) points, the second entry stores the average of the next $m$ points, and so on. Similarly, each entry at level $l$ is the average of $m$ entries at level $l-1$. As such, stored data represent local averages of the original time series and the averaging window size ($m^l$ for level $l$) increases with the level, so does the time lag it covers (the range of previous time where information is retained at the current level). For the smallest time lags (up to $p-1$ sampling intervals), the unfiltered time series is used, whereas for increasing time lag (higher levels), averages over larger window sizes are used. In this study, we use the same default parameters $p=16$ and $m=2$ as recommended in \citet{Ramirez2010}.

\begin{figure}
		\centering
		\begin{subfigure}[b]{3.5in}
			\centering
			\includegraphics[width=\textwidth]{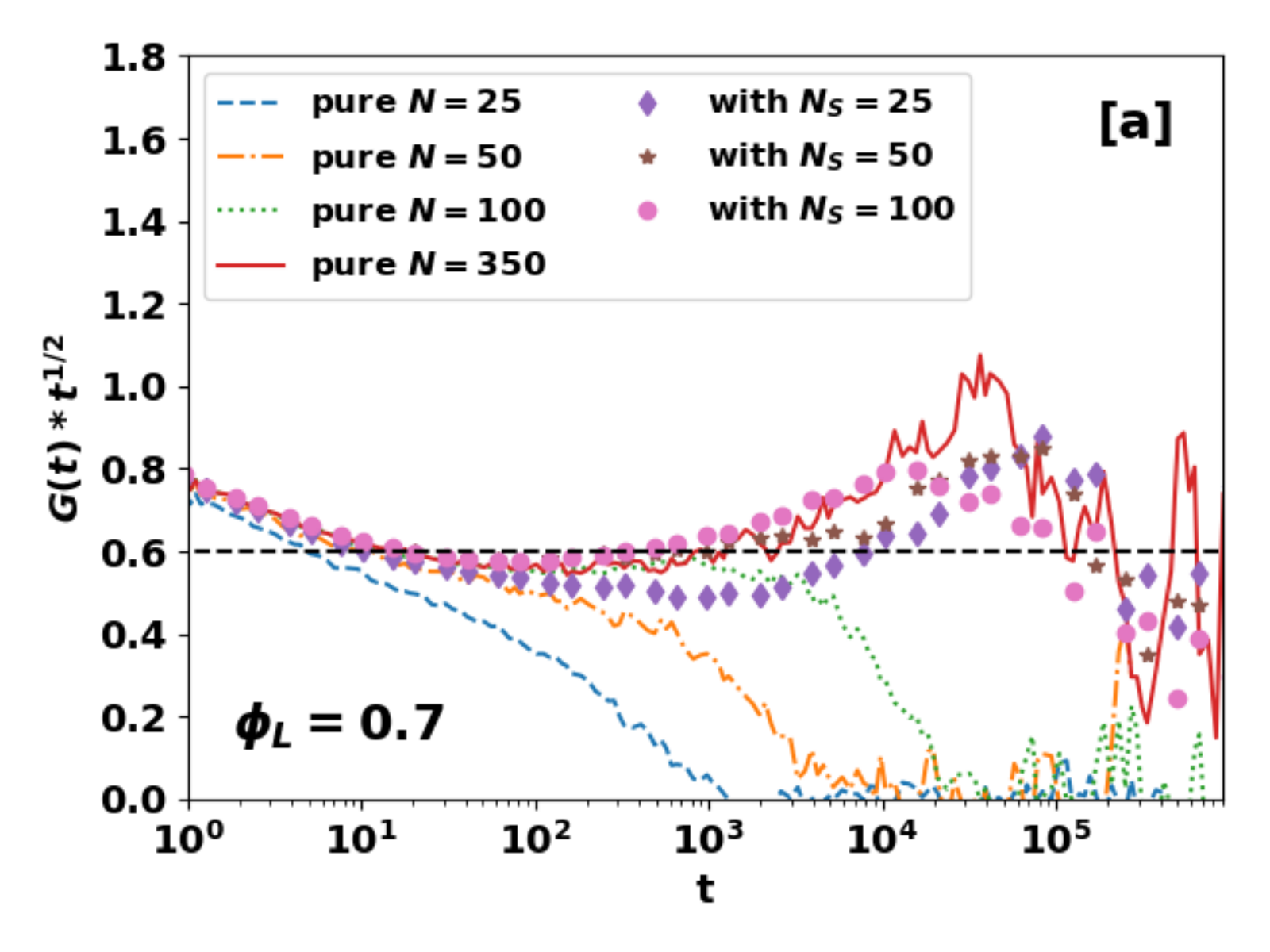}
			\phantomsubcaption
			\label{fig:Gt_bidisperse_350_major}
		\end{subfigure}
		\begin{subfigure}[b]{3.5in}
			\centering
			\includegraphics[width=\textwidth]{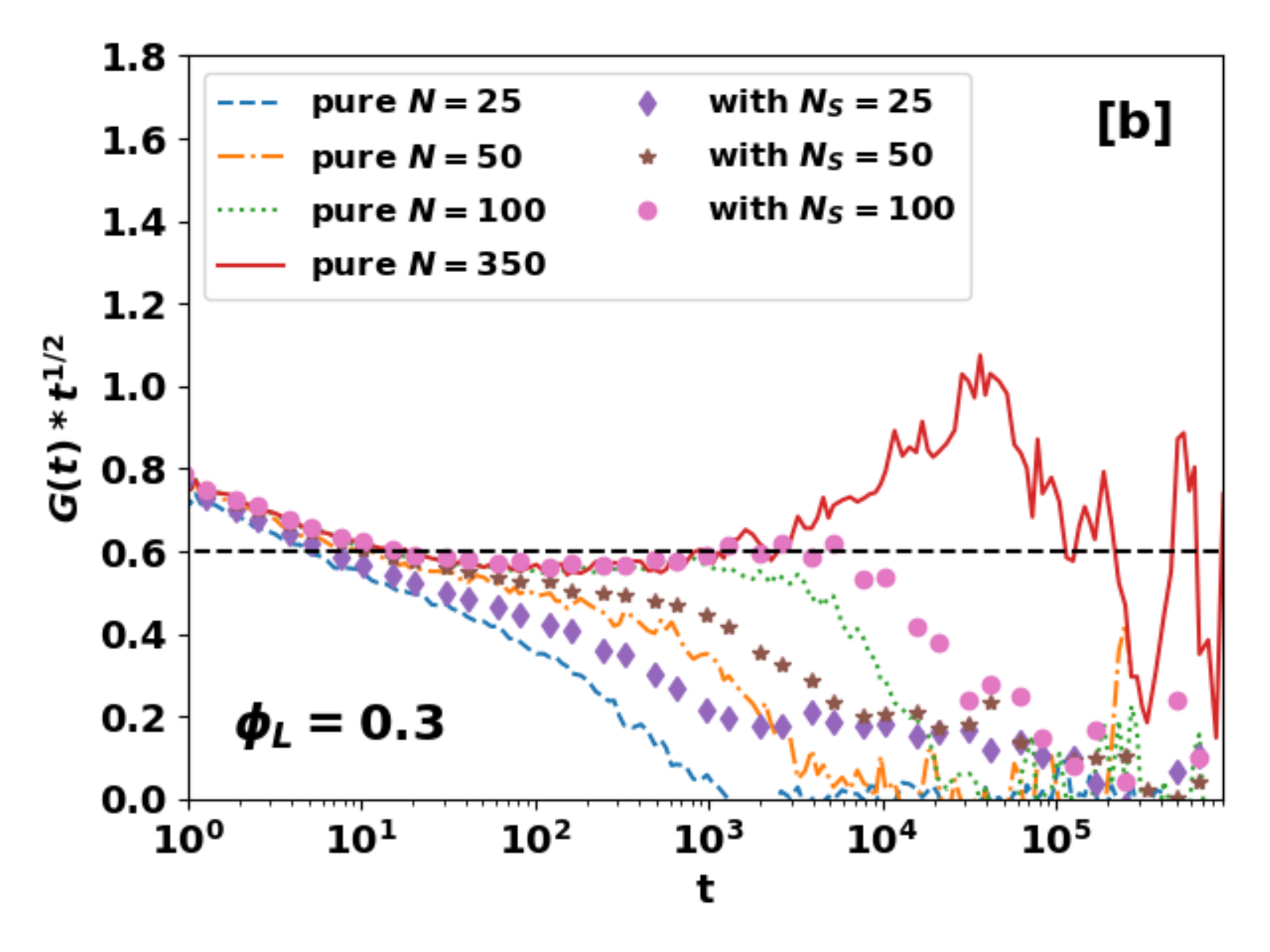}
			\phantomsubcaption
			\label{fig:Gt_bidisperse_350_minor}
		\end{subfigure}
		\caption{Stress relaxation modulus $G(t)$ (scaled by $t^{1/2}$) of pure polymer melts of different chain length $N$ (lines) and bidisperse melts with $N_\text{L}=350$ and various $N_\text{S}$ (symbols): (a) $\phi_\text{L}=0.7$\RevisedText{;} (b) $\phi_\text{L}=0.3$. Each profile is averaged over three independent configurations. Horizontal dashed line shows the Rouse model prediction.}
		\label{fig:Gt_bidisperse}
	\end{figure}

	Relaxation modulus is calculated for both monodisperse and bidisperse samples in our simulations. In \cref{fig:Gt_bidisperse}, $G(t)$ is scaled by a factor of $t^{1/2}$, as the Rouse model predicts a $t^{-1/2}$ scaling in stress relaxation: 
	\begin{gather}\label{eq:Rouse_Gt}
		G^\text{Rouse}(t) =\frac{\sqrt\pi}{2\sqrt 2}\nu_\text{chain}k_\text{B}T\left(\frac{t}{\tau_\text{R}}\right)^{-\frac{1}{2}}
		\quad (t\lesssim\tau_\text{R})
	\end{gather} 
	where $\nu_\text{chain}$ is the number density of chains, related to the bead number density $\nu_\text{bead}$ by
	\begin{gather}
		\nu_\text{chain}=\frac{\nu_\text{bead}}{N}.
	\end{gather}
	Since $\tau_\text{R}$ scales with $N^2$,
	for given monomer species and bead density, $G(t)$ magnitude does not depend on the chain length -- chain length only affects $\tau_\text{R}$, i.e., when terminal relaxation (departure from \cref{eq:Rouse_Gt}) kicks in. Using $\nu_\text{bead}=0.85$ and
	$\tau_\text{R}=1.66\times10^5$ based on the MSD of the pure $N=350$ case from our simulation (\cref{tab:timescales}), the Rouse prediction of $t^{1/2}G^\text{Rouse}(t)=0.62$ (in LJ reduced units with $G$ nondimensionalized by $k_\text{B}T$) is shown as a flat dashed line in \cref{fig:Gt_bidisperse}.
	
	It is observed that pure melts of the shorter ($N=25$ and $50$) chains completely relax before a pronounced Rouse plateau can be formed. For the pure $N=350$ melt, a Rouse plateau is clearly identified from $t\sim\mathcal{O}(10)$ up to $t\approx 2000$. For comparison, $\tau_\text{e}=3428$ according to \cref{tab:timescales}. Departure from the Rouse plateau at the small $t$ limit is also consistent with the earlier conclusion from RMA that the Rouse model does not accurately capture the dynamics at small scales. Entanglement manifests as a strong spike above the Rouse plateau at longer time. For $N=100$, entanglement is not strong enough to cause substantially raised stress than the pure rouse level. For mixtures between $N_\text{L}=350$ chains with a shorter component, if the long chains remain the majority (fig. \ref{fig:Gt_bidisperse_350_major}), a spike is still clearly observable for different diluent chain length (down to $N_\text{S}=25$). If the long chains become the minority (fig. \ref{fig:Gt_bidisperse_350_minor}), $t^{1/2}G(t)$ no longer rises above the Rouse plateau. However, compared with the short-chain cases, in bidisperse mixtures, a second lower ($\RevisedText{\sim}\,0.2$) plateau regime is found, indicating that even the weak entanglement present in those mixtures can appreciably hinder relaxation at longer time scales.

	In practice, contributions of different components of a polydisperse mixture are usually accounted for through mixing rules that relate the MWD of the polydisperse mixture to its $G(t)$. A generalized form of the mixing rule is given as\cite{anderssen1998theoretical,maier1998evaluation,tuminello1999determining}
	 \begin{equation}\label{eq:mixing_rule}
	 	G(t) = {G_N^0}{\left(\int_{\log N_\text{e}}^{\infty}F^{1/\beta}(t,N)w(N)d(\log N)\right)}^\beta 
	 \end{equation}
	where ${G_N^0}$ is the plateau modulus
	and $F(t,N)$ is a kernel function
	accounting for the contribution from chains of length $N$.
	The $w(M)$ function is given by
	\begin{equation}
		w(N)\equiv\frac{dW(N)}{d\text{log}N}
	\end{equation}
	where $W(N)$ is the cumulative weight fraction of chains with the degree of polymerization lower than $N$, and $\beta$ is a parameter describing the mixing behavior. Different $\beta$ values have been used in the literature.
	A $\beta$ value of $1$ gives linear combination derived from the simple tube model.
	The double reptation model by des Cloizeaux \citep{des1992relaxation} leads to a $\beta$ value of 2. \citet{VanRuymbeke2002} adjusted the value to 2.25 to obtain better fits for the linear viscoelastic properties evaluated in their study. Higher $\beta$ values have also been proposed in the literature to describe higher order entanglements \cite{VanRuymbeke2002,marrucci1985relaxation,thimm2000rouse}. Several kernel functions have also been reported.
	The simplest choice is a step function\cite{tuminello1986molecular} which assumes
	steep transition between strained and relaxed conformations.
	\citet{tsenoglou1991molecular} described the relaxation using a more realistic single exponential function which gives better qualitative prediction. More accurate quantitative description is possible with more sophisticated forms such as the relaxation function of \citet{doi1988theory}, along with its derived form that accounts for the effects of contour length fluctuations\cite{doi1988theory}, and the time-dependent diffusion model of \citet{des1990relaxation}.

	In a bidisperse mixture,
	\begin{equation}
		w(N)=\phi_\text{S}\delta(\log N - \log N_\text{S})+\phi_\text{L}\delta(\log N - \log N_\text{L}).
	\end{equation}
	Using the property
	\begin{equation}
		\int_{a-\epsilon}^{a+\epsilon}f(x)\delta(x-a)dx=f(a) \quad (\epsilon >0)
	\end{equation}
	of the Dirac delta function,
	\cref{eq:mixing_rule} becomes
	\begin{equation}\label{eq:simplified_mixing_rule}
		G_\text{S+L}(t)={G_N^0}\left(\phi_\text{S}F_\text{S}^{1/\beta}(t)+\phi_\text{L}F_\text{L}^{1/\beta}(t)\right)^\beta
	\end{equation}
	where
	\begin{gather}
		F_\text{S}(t)\equiv F(t,N_\text{S})
	\end{gather}
	and
	\begin{gather}
		F_\text{L}(t)\equiv F(t,N_\text{L})
	\end{gather}
	are the kernel functions of the short- and long-chain species, respectively.

	Our focus here is not on the analytical theory of the relaxation of individual chain species itself, but on predicting mixture rheology based on the relaxation behaviors of individual components. Therefore, we circumvent the analytical expression of the kernel function and extract it directly from MD simulation of corresponding monodisperse melts. Note that at the $\phi_\text{S}\to1$ limit, \cref{eq:simplified_mixing_rule} becomes
	\begin{gather} \label{eq:Gs}
		G_\text{S}=G_N^0 F_\text{S}(t)
	\end{gather}
	and at the $\phi_\text{L}\to1$ limit
	\begin{gather} \label{eq:Gl}
		G_\text{L}=G_N^0 F_\text{L}(t).
	\end{gather}
	Combining \cref{eq:Gs,eq:Gl} with \cref{eq:simplified_mixing_rule} leads to
	\begin{gather}
		G_\text{S+L}=\left(\phi_\text{S}G_\text{S}^{1/\beta}(t)+\phi_\text{L}G_\text{L}^{1/\beta}(t)\right)^\beta
		\label{eq:mixingrulebinary}
	\end{gather}
	where $G_\text{S}(t)$ and $G_\text{L}(t)$ are obtained from the corresponding monodisperse simulation results.
	
	\begin{figure}
		\centering
		\begin{subfigure}[b]{3.5in}
			\centering
			\includegraphics[width=\textwidth]{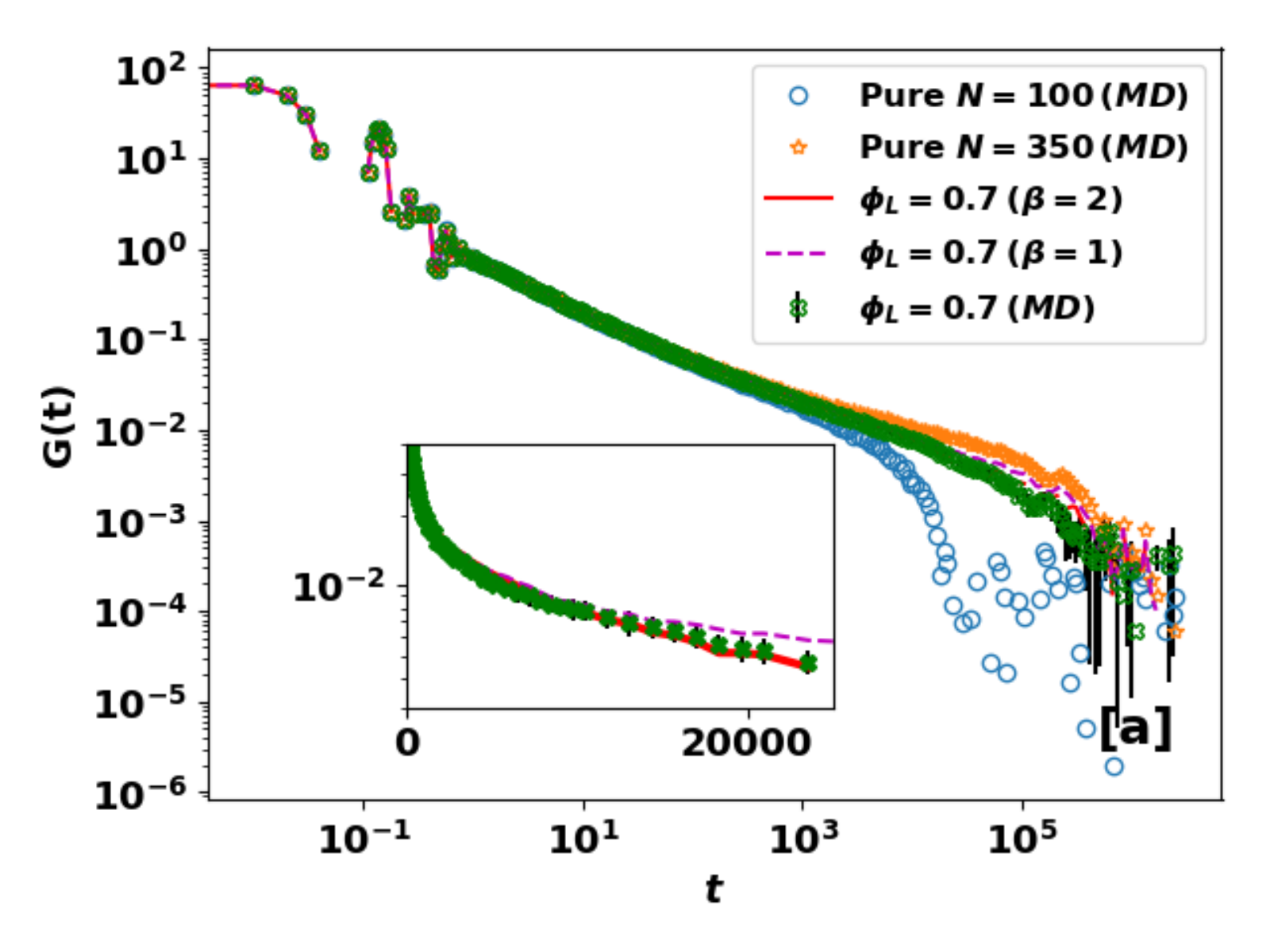}
			\phantomsubcaption
			\label{fig:Mixing_Rule_350_100}	
		\end{subfigure}		
		\begin{subfigure}[b]{3.5in}
			\centering
			\includegraphics[width=\textwidth]{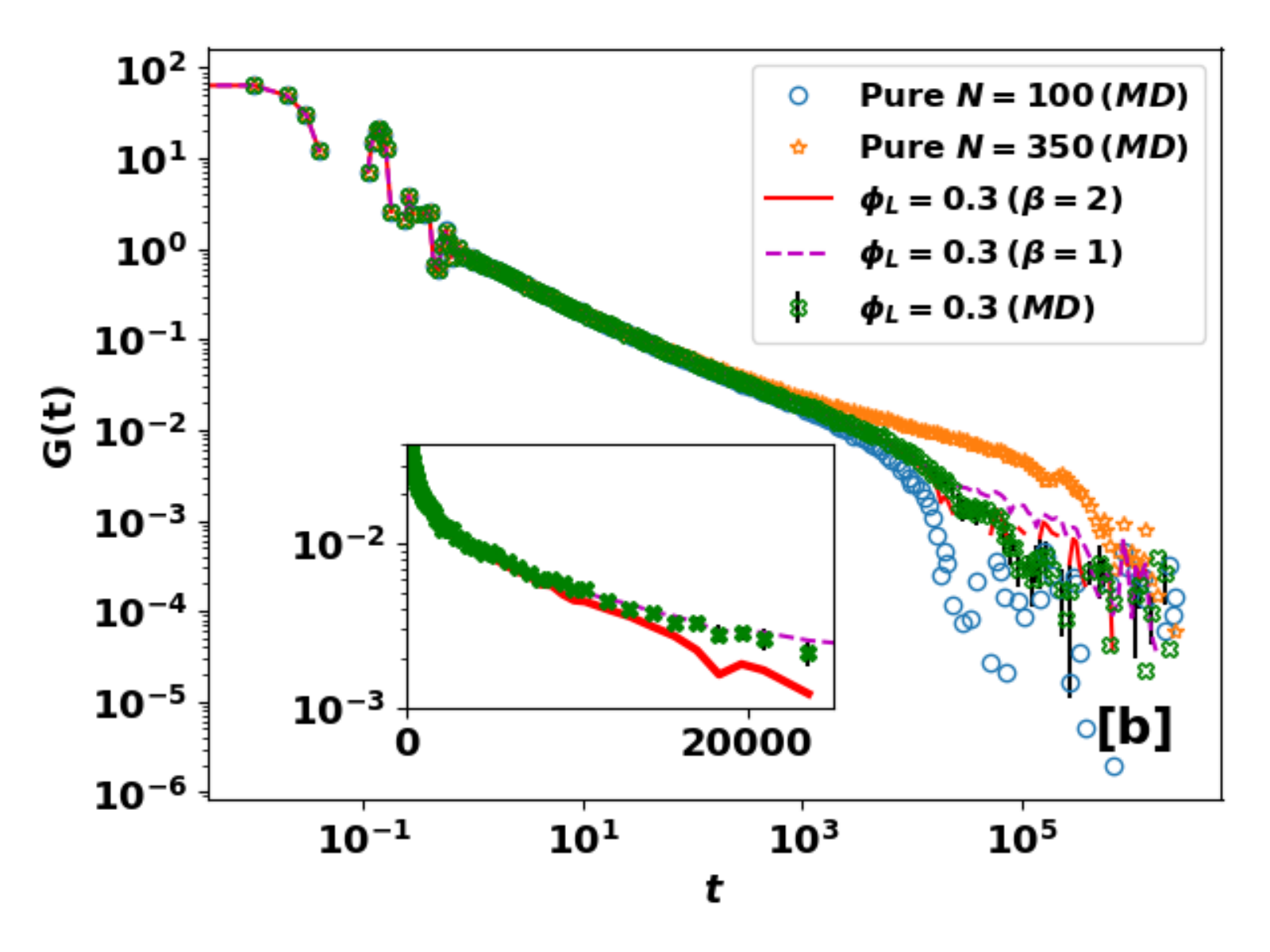}
			\phantomsubcaption 
			\label{fig:Mixing_Rule_100_350}
		\end{subfigure}
		\caption{Test of the mixing rule (\cref{eq:mixing_rule}) in bidisperse mixtures of $N_\text{S}=100$ and $N_\text{L}=350$ at (a) $\phi_\text{L}=0.7$ and (b) $\phi_\text{L}=0.3$.
		Insets show enlarged views of the comparison between the mixing rule and MD before terminal relaxation in a linear time scale.
		\RevisedText{Error bars are shown for bidisperse MD cases only, which are no larger than the symbol size except in the terminal-relaxation (large-$t$) regime.
		Uncertainties in pure-melt MD cases are similar.}} 
		\label{fig:Mixing_Rule}
	\end{figure}  

	Predictions from \cref{eq:mixingrulebinary} are shown in \cref{fig:Mixing_Rule} in comparison with direct MD results of bidisperse mixtures. Since the mixing rule of \cref{eq:mixing_rule} is only applicable to entangled melts, binary mixtures between $N_\text{S}=100$ and $N_\text{L}=350$ are selected here. Corresponding monodisperse MD results are also displayed.
	At early times, a regime dominated by Rouse relaxation, all the curves overlap as expected. Per discussion above, $G(t)$ from the Rouse model is independent of chain length at $t<\tau_\text{R}$. At longer times, dynamics of different cases separate with the pure $N=100$ case being the first to decay and $N=350$ being the last. Relaxation dynamics of bidisperse mixtures is sandwiched between the two monodisperse limits.
	
	Comparing the mixing rule results with MD of bidisperse cases, $\beta=2$ gives strikingly accurate prediction at $\phi_\text{L} = 0.7$ as shown in \cref{fig:Mixing_Rule_350_100}, showing that the double reptation model predicts the relaxation of the system reasonably well.
	\RevisedText{Note that for $t$ up to at least $\mathcal{O}(\num{e4})$, error bars in the MD data are smaller than the symbol size and thus the comparison is statistically significant.}
	For $\phi_\text{L} = 0.3$, $\beta = 1$ seems to be more accurate for $t\lesssim\num{2e4}$ shown in the inset of \cref{fig:Mixing_Rule_100_350}, but
	\RevisedText{as we examine $t>\num{2e4}$, we spot a kink in the MD profile, which brings the curve closer to the $\beta=2$ line right before terminal relaxation.
	Admittedly, fluctuations (and statistical uncertainty) also grow in that regime as we move closer to terminal relaxation, making the observation less statistically conclusive than the earlier agreement with $\beta=1$ at smaller $t$.}
	We have also tested $\beta = 2.25$ which gave worse results and is thus not shown here.
	\RevisedText{At this point, we are not ready to interpret these observations, including (i) the concentration-dependence of the $\beta$ value and (ii) \emph{possible} switch from $\beta=1$ to $\beta=2$ at later time in the $\phi_\text{L}=0.3$ case.
	Definite answers will require expansive simulations including multiple concentration levels and a wider range of chain lengths.
	}
	
\section{Conclusions}
We have studied the chain dynamics and stress relaxation of bidisperse polymer melts using MD simulation. For each bidisperse system, we mixed a long $N_\text{L}=350$ chain component, whose monodisperse melt is entangled, with a short-chain diluent (which is either unentangled -- i.e., $N_\text{S}=25$ and $50$, or marginally entangled $N_\text{S}=100$). 
Two different composition levels, one with the long chains as the majority component ($\phi_\text{L}=0.7$) and the other as the minority ($\phi_\text{L} = 0.3$) component, were studied.

Compared with a pure short-chain melt, mixing with longer chains significantly reduces the mobility of short chains. However, dynamics of a short, unentangled, chain species remains well-described by the Rouse model, despite the presence of the longer, entangled, chain species in the mixture.
At least for $\phi_\text{S}$ down to $30\%$ studied, dynamics of a short chain in a slow-moving matrix containing entangled long chains shows the same qualitative pattern as its relaxation in a pure monodisperse melt.
Slow-down in the dynamics can be well captured by a higher effective monomeric friction coefficient.

Likewise, adding a short-chain diluent can significantly accelerate the motion of the longer, entangled, chain species. 
The effect is stronger as the short-chain mass fraction increases and as its chain length decreases.
Unlike the previous case, however, this speed-up effect cannot be fully described by a quantitatively lower friction coefficient, which is instead accompanied by an overall lessening of the extent of entanglement.

Rouse mode analysis revealed that for short, unentangled, chains, relaxation of Rouse modes displays simple exponential decay, except the highest modes corresponding to dynamics at smallest scales.
Mixing with longer chains does not change this qualitative observation.
Meanwhile, for longer, entangled, chains, strong departure from simple-exponential relaxation is seen at intermediate length scales of $\mathcal{O}(N_\text{e})$.
Relaxation times of intermediate and large scales also become significantly elevated compared with the Rouse model.
Both these characteristics of entanglement become weakened with the introduction of short-chain diluents.
	
Despite the success of the Rouse model in describing the dynamics of short chains (in both monodisperse melts and bidisperse mixtures with longer chains), stress relaxation of monodisperse short chains does not strictly follow the Rouse scaling.
For monodisperse entangled chains, a well-defined Rouse scaling regime is observed, followed by a positive deviation from the Rouse model at longer times (in the entanglement regime).
Bidisperse mixtures display similar positive deviation when the longer species is the majority.
When it becomes the minority, the relaxation modulus no longer surpasses the Rouse scaling. Entanglement is instead reflected as an elongated (compared with a pure short-chain melt) tail of residue modulus.

Mixing rules for predicting a mixture's relaxation modulus from that of monodisperse melts of its constituent components were tested.
The double reptation model provides reasonably accurate prediction when the longer chains are the majority. However, when longer chains are the minority, the simple tube model can be more accurate in certain regimes.

\begin{acknowledgments}
The authors acknowledge the financial support from the Natural Sciences and Engineering Research Council of Canada (NSERC) Discovery Grants program (No.~RGPIN-2014-04903 and No.~RGPIN-2020-06774) and the allocation of computing resources awarded by Compute/Calcul Canada.
S.Z. thanks the Canada Research Chairs (CRC) program (No.~950-229035).
This work is made possible by the facilities of the Shared Hierarchical Academic Research Computing Network (SHARCNET: www.sharcnet.ca).
\end{acknowledgments}

\RevisedText{%
\section*{Dedication}
This paper is dedicated to Robert Byron Bird (1924--2020) whose extraordinary contributions to the research and education in the dynamics and rheology of polymers, kinetic theory of fluids, and transport phenomena were matched by very few.
His love of cultures, sense of humor, and passion for life continue to inspire many.}%
	
\section*{Data Availability}
The data that support the findings of this study are available from the corresponding author upon reasonable request.
	
\bibliography{PolymerDynamics}

\end{document}